\documentclass[12pt]{article}

\usepackage{eurosym,geometry,ulem,color,setspace,sectsty,comment,footmisc,caption,pdflscape,subfigure,array}

\usepackage{amsmath,amssymb,amsfonts}
\usepackage{graphicx}
\usepackage{textcomp}
\usepackage{xcolor}
\usepackage{booktabs}
\usepackage{hyperref}
\usepackage{url}
\usepackage{nicefrac}
\usepackage{microtype}
\usepackage{bm}
\usepackage[ruled,linesnumbered]{algorithm2e}
\usepackage{mathtools}
\usepackage{multirow}
\usepackage{lscape}

\usepackage{natbib}
 \bibpunct[, ]{(}{)}{,}{a}{}{,}%
 \def\BIBand{and}%

\newcolumntype{L}[1]{>{\raggedright\let\newline\\arraybackslash\hspace{0pt}}m{#1}}
\newcolumntype{C}[1]{>{\centering\let\newline\\arraybackslash\hspace{0pt}}m{#1}}
\newcolumntype{R}[1]{>{\raggedleft\let\newline\\arraybackslash\hspace{0pt}}m{#1}}

\begin{document}

\begin{titlepage}
\title{Neural Learning of Online Consumer Credit Risk}
\author{
	Di WANG \\ JD Digits \\ albertwang0921@gmail.com
	\and
	Qi WU\thanks{The corresponding author.}\\City University of Hong Kong \\ qiwu55@cityu.edu.hk
	\and
	Wen ZHANG\\JD Digits\\ zhangwen11@jd.com
}
\date{May 31, 2019}
\maketitle
\begin{abstract}
\noindent This paper takes a deep learning approach to understand consumer credit risk when e-commerce platforms issue unsecured credit to finance customers' purchase. The ``NeuCredit" model can capture both serial dependences in multi-dimensional time series data when event frequencies in each dimension differ. It also captures nonlinear cross-sectional interactions among different time-evolving features. Also, the predicted default probability is designed to be interpretable such that risks can be decomposed into three components: the subjective risk indicating the consumers’ willingness to repay, the objective risk indicating their ability to repay, and the behavioral risk indicating consumers' behavioral differences. Using a unique dataset from one of the largest global e-commerce platforms, we show that the inclusion of shopping behavioral data, besides conventional payment records, requires a deep learning approach to extract the information content of these data, which turns out significantly enhancing forecasting performance than the traditional machine learning methods.\\
\vspace{0in}

\bigskip
\end{abstract}
\setcounter{page}{0}
\thispagestyle{empty}
\end{titlepage}
\pagebreak \newpage

\doublespacing

\section{Introduction} \label{sec:introduction}
\begin{figure}[t]
	\centering
	\includegraphics[scale=0.6]{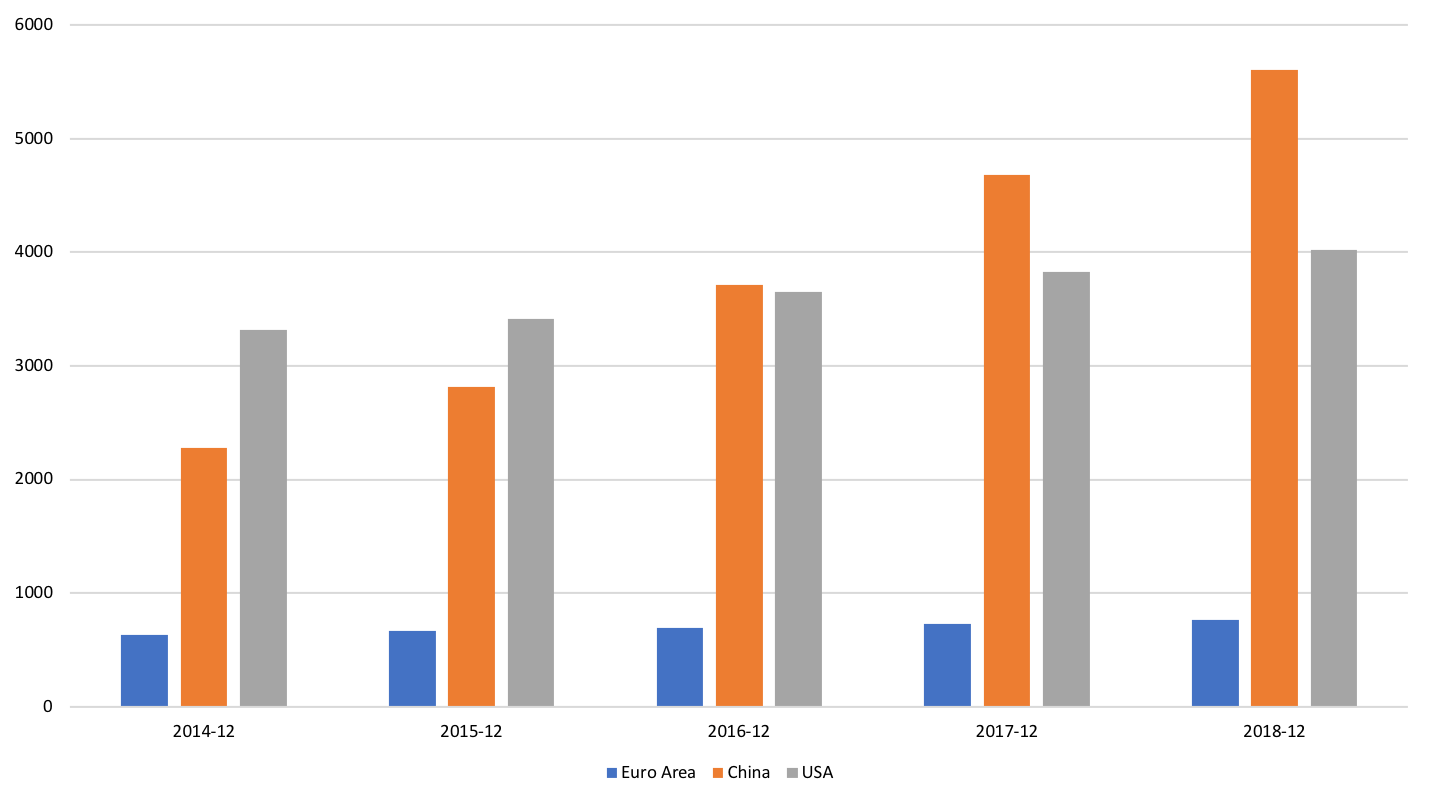}
	\caption{Consumer credit outstanding across years and countries/areas. The horizontal axis indicates the time. The amounts are the stocks at the end of each period. The vertical axis measures the amount of outstanding consumer credit. All currencies are denominated in US dollar. The colors suggest different entities. The Eurozone, China, and the United States are marked blue, orange, and gray, respectively. Data sources: European Central Bank, National Bureau of Statistics of China, and Federal Reserve.}
	\label{fig:market}
\end{figure}

The consumer credit market in the Eurozone, the United States, and China went up dramatically since 2014. According to the European Central Bank, the US Federal Reserve, and the National Bureau of Statistics of China\footnote{The data is from official releases. US: Federal Reserve (www.federalreserve.gov/Releases/G19/current/). Euro Area: European Central Bank (www.euro-area-statistics.org/banks-balance-sheet-loans). China: National Bureau of Statistics of China (www.stats.gov.cn/tjsj/)}, the outstanding notional at the end of 2018 is 770, 4,018 and 5608 billion US dollars, respectively, with China being the most notable in terms of the size of the market and the speed of the growth (see figure \ref{fig:market}). While property financing such as housing and automobile remains the main driver, a fast-growing portion comes from people's spending on credit for necessities and consumables. One reason is that technology enables credit to channel into greater coverage of population and deeper penetration of consumer spending. A case in point is the credit issuance through global e-commerce platforms.  Tremendous purchasing and borrowing activities now migrate from offline to online. For researchers, this paradigm shift from offline to online opens the door to observe consumer behavior at an unprecedented granularity, presenting new opportunities to decipher retail credit risk, and at the same time, new challenges to credit risk modeling. 

Retail credit risk is the risk of capital loss when consumers fail on payments of credit card or personal loan. Traditionally, analysis of consumer credit risk focuses on credit score using low-frequency data where maintaining a good payment record play a dominant role. In these analyses, characteristics regarding customer's purchasing activities are either not available or not included. Whether it is a teenager buying a ten thousand dollar watch or it is a business owner buying a laptop, their credit scores are likely not very different to a credit card company as long as they pay on time. However, in the e-commerce context, consumers' shopping footprints and the subsequent purchasing activity are naturally connected with their credit-seeking and payment records. Including these behavioral data into the credit analysis allows online risk managers to tell one from another both their willingness to repay and their ability to repay with better confidence. 

A typical cycle of online shopping on credit consists of three stages of actions. A customer first browses items that she is interested in. She then places orders on items she decides to buy and starts to think whether and how much payment credit she would like to apply from the platform. Once the credit is granted if she did apply for it, she effectively enters into an unsecured loan with the platform as the lending counterparty, and she is expected to make installments according to the payment schedule. These decisions and events, which we classify into three distinct groups (\textit{browsing}, \textit{ordering}, and \textit{borrowing}), collaboratively shape the credit profile of a customer. As these decisions and events intrinsically indicate a consumer's ability and the willingness to repay, modeling credit risk based on them seems promising. However, the behavioral nature of these data and the granularity they present pose several daunting challenges. 

First, \textit{the three groups of actions take place very irregularly in time}. This is due to their wide spectrum of event frequencies, therefore causing distinct degrees of serial-dependencies. The event frequencies can range from hundreds of times a day such as browsing activities, to once in a few days where subsequent purchases occur, and all the way to quarterly or semi-annual frequencies when periodical installments on borrowing are either paid or past due. This makes it difficult to learn the temporal dynamics of consumer behavior if one wants to use all three group of data together. This irregularity can also cause serious dominant view problem in model fitting. For example, browsing happens much more frequent than the other two groups of actions. In this situation, browsing data easily dominates the feature space, especially over those events that are very informative yet occurred much less often, such as defaults.

Second, \textit{the three groups of actions are interacting with each other in a complex manner}. The relationship between browsing, ordering, and borrowing activities can be highly nonlinear. For instance, an increase of browsing activities may result in more purchases if the costumer's financial wellbeing is healthy, but it may also cause less if the customer realizes that her accumulated spending in the past is about to reach her financial limit and thus becomes more cautious. The three groups of actions have their own information heterogeneity and complement each other in reflecting the behavioral pattern of a customer. It is a challenge to model these complex interactions in time series effectively. 

Third, \textit{how to interpret the predicted result regarding a consumer's credit risk is critical.} In other words, finding the determinants are as important as predicting the outcome from the perspective of credit risk management. While enlarging datasets to include rich behavioral information surely leads to better model estimation and more accurate forecasts, the complexity of interpreting results also increases. Nevertheless, carefully exploiting the browsing and ordering actions as well as the outcomes of related borrowing should shed lights on whether a customer is going to default or not and if so, why. 

This paper develops a deep neural network (DNN) model to estimate and forecast consumer credit risk, and at the same time provide a structural attribution of the perceived risk into a consumer's ability to repay factor, her willingness to repay factor, and her behavioral factor. We call it the \textit{NeuCredit} model and test it on a unique data set collected from one of the largest global e-commerce platforms. The dataset contains real-world proprietary records collected by one of the largest global e-commerce platforms. It includes 38,182 loans with 499,572 relevant orders and 356,338 relevant sessions of clicks. The goal is to estimate the real-time default risk when a customer uses her approved credit to finance a purchase.

In particular, the model features a \textit{hierarchical architecture} in which three groups of actions are processed separately to avoid the problem of the dominant view. The sequence of borrowing actions that specifies the time-stamps of loan issuance is regarded as the mainstream, i.e., the first layer, while the browsing and ordering actions are respectively clustered to their nearest future loan to form two sub-sequences (the second layer) for each loan. Considering the sequential nature of data, we propose a variant of Long Short Term Memory (LSTM) model, named the \textit{Time-value-aware LSTM} (Tva-LSTM) model, to learn the temporal dynamics of irregular consumer behavior. By assuming the effect of an action in future prediction is continuously growing or decaying at trainable rates, the Tva-LSTM model captures the varying time intervals between every two consecutive actions in time series. Furthermore, sub-sequences are integrated into the mainstream through a novel \textit{multi-view fusion} mechanism that explicitly models the mutual effects via feature interactions. The fusion is performed in nearly real-time as it launches at each element of the mainstream. We supervise the training of the NeuCredit model using labeled data, i.e., whether a consumer is delinquent or she defaults on her payments. 

We conducted extensive experiments to validate the effectiveness of the NeuCredit model, followed by regressions to understand the learning result. Comparing with conventional and other state-of-the-art models, the NeuCredit model successfully captures the complex behavioral dynamics and improves the performance of consumer credit risk estimation. It achieves remarkable performance not seen before in out-of-sample forecasts. In particular, the model can capture both serial dependences in multi-dimensional time series data when event frequencies in each dimension differ. It also captures nonlinear cross-sectional interactions among different time-evolving features. Besides, the predicted credit risk is designed to be interpretable such that risks can be decomposed into three components: the subjective risk indicating the consumer's the willingness to repay, the objective risk indicating their ability to repay, and the behavioral risk indicating their behavioral differences. The willingness and the ability of customer repaying are modeled into the neural network via a specially-designed conditional loss function even though their ground-truths are unobservable.

\subsection{Our Contribution}
The contributions as well as messages of this study are threefolds.

\begin{itemize}
    \item  {\textbf{Tick-level shopping behavioral data enhances online credit risk forecasts.}} 
    
    The underlying relationship between consumer shopping behavior and their credit risks has not been formally studied before. In this paper, we profile consumer credit at an unprecedented granular level by zooming into the tick-level shopping behavior and the subsequent financing records. Deciphering them carefully allows real-time assessment of future payment risk, particularly when online purchases are financed without posting collateral. Our extensive experiments demonstrate that online credit risk forecasts are improved significantly when browsing and purchasing data are added into the model training, comparing to using only the payment data. To the best of our knowledge, this is the first academic study that focuses on consumer credit risk in e-commerce contexts using a large comprehensive dataset to model consumer delinquencies and defaults.
    
    \item {\textbf{Deep learning approach outperforms conventional machine learning  significantly.}} 
    
    We propose a novel LSTM-based deep learning framework designed to handle complex consumer behavior, especially the irregularities of sequential actions and the interactions across different groups of actions. Here, we propose a hierarchical network structure and a Tva-LSTM unit to handle temporal sequences of irregular consumer actions. Besides, we design a multi-view fusion mechanism to model action interactions so that it can uncover the mutual effects of different groups of shopping behavior. The model is effective: empirical results demonstrated its performance superiority over the conventional machine learning model such as the logistic regression model and the random forest model as well as the competing state-of-the-art deep learning models using the LSTM architecture. To the best of our knowledge, this is the first systematic study of consumer credit risk modeling using an LSTM-based deep learning approach. Moreover, the framework is generic in that one can use it to in non-financial applications such as recommendation, anomaly detection, etc. The source codes of our algorithm are available upon request.
    
    \item {\textbf{Our model outputs structural interpretation of the risk determinants.}} 
    
    The deep learning models are often criticized for their black-box nature and the lack of interpretability. Our approach to addressing this issue is to propose a specially-designed conditional loss objective in order to incorporate domain knowledge into the system. Specifically, the ability and the willingness to repay are considered as two significant loan determinants defaults \citep{lee1991ability,chehrazi2015dynamic} in credit risk management. Understanding their contributions to the predicted credit risks is, therefore, informative. It helps a risk manager to identify the sources of credit risk and makes informed decisions on debt collection and credit extension. However, as ability and willingness cannot be observed in consumer actions directly, their ground-truths are not available in modeling. Here, we inferred their values through the repaying outcomes of loans and designed a conditional loss function to take these inferred values as guidance. In this way, the system can generate interpretable outputs. In the literature, this is the first deep learning approach that provides interpretable predictions of consumer credit risk.
\end{itemize}

The rest of the paper is organized as follows. We first review related works in Section \ref{sec:literature}. We then give descriptions of the dataset we use in Section \ref{sec:data}. Section \ref{sec:method} introduces our model. Experiments are presented and analyzed in Section \ref{sec:experiment}. We conclude the paper in Section \ref{sec:conclusion} together with a discussion on possible future directions.

\section{Literature Review} \label{sec:literature}
Our paper is related to the machine learning approach to the modeling and understanding of consumer credit risk. Academic studies concerning retail credit are fewer comparing to the vast majority of the credit risk literature that is corporate, sovereign or mortgage oriented. One reason is that there is little outright trading of individual personal loans, hence no public assessments of retail credit risk. Unlike corporate bonds, secondary trading of securities related to consumer credit are only in secularized form\footnote{In US market, Asset-Back-Securities backed up with credit-card proceeds are liquid.}. Another reason is the lack of account-level data unless one has access to proprietary data owned by commercial banks and credit card companies. In terms of risk metrics and the models used, the historical focuses are credit scoring and linear regression when it comes to consumer credit risk. However, as e-commerce plays an ever-larger role in retail credit insurance and much richer data becomes available, sophisticated credit models are needed for the management of retail credit risk.

Earlier work using machine learning approach to analyze consumer credit-risk starts from Khandani, Kim, and Lo (\citeyear{khandani2010consumer}) where classification and regression trees are used to construct forecasting models. Using a unique dataset consisting of transaction-level, credit bureau and account-balance data for individual consumers, they were able to forecast credit events related to consumer credit default and delinquency 3-12 months in advance with great accuracy. The results in Khandani, Kim, and Lo (\citeyear{khandani2010consumer}) show that machine learning approach is very suitable to build forecasting models when the sources of information are vast, the nature of data is distinct, and the connections between them are unclear.

Sirignano, Sadhwani, and Giesecke (\citeyear{sirignano2016deep}) advance the machine learning approach to credit risk modeling from classical machine learning methods to deep neural networks. Comparing to classical machine learning models, the recurrent neural networks (RNN) used in Sirignano, Sadhwani, and Giesecke (\citeyear{sirignano2016deep}) are extremely capable of extracting nonlinear relationships between explanatory variables and response variables. These nonlinear relationships are shown to be very important in the out-of-sample forecast when benchmarked with linear models such as logistic regression.  Using a dataset of over 120 million mortgages and over 3.5 billion loan-month observations across the US between 1995 and 2014, the authors demonstrate the powerfulness of RNN in terms of estimating transition probabilities of credit states and understanding of mortgage credit and prepayment risk at an unprecedented level. 

Our paper further adds to the literature on using a machine learning approach to study consumer credit. Methodology-wise, the first comparative merit of our model is its interpretability. The neural network architecture we design can output interpretable factors in order to understand what drives the consumer defaults and delinquencies, such as "the willingness to repay" factor and "the ability to repay" factor suggested earlier in the literature \citet{lee1991ability}.

The second merit of our model is its ability to allow irregular time interval in data when learning complex serial dependence in high dimensional time series. Our findings coincide with the study of \citet{chehrazi2015dynamic} where self- and cross-excited Hawkes process captures dependencies between the arrival times of repayment events. The authors show that it is essential to capture the dependence structure when account-level data is used either for valuation or forecasting. Since our data show a wide spectrum of event frequencies, ranging from hundreds of times a day in browsing activates all the way to monthly or quarterly frequencies in payment installments, we need more flexibility than previous machine learning approaches to model potentially distinct degrees of serial-dependencies and complex nonlinear cross-sectional interactions. 
Thus, the deep neural network we construct uses a hierarchical architecture rather than outright RNN or classic machine learning methods. On top of that, the LSTM specification we use addresses the issue that traditional RNN is not very good at learning long term memories in the data, but keeping the nonlinear mapping ability of RNN between inputs and outputs.

\section{Data Description} \label{sec:data}
The data set is from one of the largest global e-commerce platforms in which, the whole courses of customers' online shopping on credit are recorded, i.e., browsing items, placing orders, seeking credit, and repaying loans. The browsing, ordering, and borrowing activities are recorded in the forms of sessions of clicks, orders, loans, respectively. 

A session of clicks is defined as beginning with a click which occurs after 15 minutes or more have elapsed since the last click and continuing until 15 minutes or more elapse between clicks. The consumers in our dataset are required to have conducted at least three times of borrowing instances on the platform during the period from Nov. 1st, 2016 to Nov. 1st, 2018, i.e., have at least three historical loans. To limit the length of loan sequence, only the most recent 15 loans of each consumer are recorded. In this way, each consumer in the dataset possesses a temporal loan sequence with a minimum length of 3 and a maximum length of 15. 

For each loan in a loan sequence, only the orders within the past 6 months before the issuance of that loan and the sessions within the past 14 days before the issuance of that loan are recorded. This is because the contribution of ordering and browsing actions in predicting default risk is considered time-sensitive. For example, it is unlikely that a customer would spend more than two weeks to make a single decision on whether to buy something. Therefore, browsing behavior more than 14 days before the current loan might not be helpful. 

Also, only the most recent 15 orders and 15 sessions before the issuance of each loan are recorded to limit the length of order and session sequences. A loan sequence that has a loan with less than 3 orders or 3 sessions before the issuance of that loan are dropped. In this way, each loan in a loan sequence possesses a temporal order sub-sequence and a temporal session sub-sequence both with a minimum length of 3 and a maximum length of 15. 
From the consumers that meet the above requirements, 2,500 of them with no default records in their loan sequence are randomly selected, and 2,500 of them with at least one default record in their loan sequence are randomly selected. A default record generates when a consumer has been delinquent for more than 90 days on a loan. In total, 5,000 consumers are selected. Finally, the dataset contains 38,182 loans where 11,184 of them are default ones, 499,572 orders, and 356,338 sessions of clicks. On average, each consumer has 7.64 loans, and each loan is related to 13.08 orders and 9.33 sessions, i.e., the average length of loan sequences, order sub-sequences, and session sub-sequences is 7.64, 13.08, and 9.33, respectively.

\begin{table}[t]
\centering
  \caption{Summary Statistics of Loans. The number of loans is 38,182, where 11,184 loans default. Variable names l.amt, term, int.rate, and l.itv represent loan amount (CNY), loan term (month), annualized interest rate (\%), and time interval between consecutive loans (day), respectively.}
  \label{tab:su_loan}
  \begin{tabular}{ccccccccc}
    \toprule
    & Variable & Mean & SD & 5th & 25th & Median & 75th & 95th \\
    \midrule
    \multirow{4}{*}{\shortstack{All\\Loans}}
    & l.amt & 322.09 & 756.71 & 18.69 & 51.49 & 107.61 & 227.01 & 1439.61 \\
    & term & 1.87 & 1.97 & 1.00 & 1.00 & 1.00 & 1.00 & 6.00 \\
    & int.rate & 2.18 & 4.41 & 0.00 & 0.00 & 0.00 & 0.00 & 12.00 \\
    & l.itv & 16.48 & 30.72 & 0.00 & 0.00 & 4.00 & 17.00 & 80.00 \\
    \midrule
    \multirow{4}{*}{\shortstack{Default\\Loans}}
    & l.amt & 312.52 & 722.52 & 19.97 & 50.01 & 105.55 & 230.83 & 1299.01 \\
    & term & 2.52 & 2.67 & 1.00 & 1.00 & 1.00 & 3.00 & 6.00 \\
    & int.rate & 3.64 & 5.26 & 0.00 & 0.00 & 0.00 & 9.60 & 12.00 \\
    & l.itv & 11.40 & 24.22 & 0.00 & 0.00 & 2.00 & 10.00 & 57.00 \\
    \midrule
    \multirow{4}{*}{\shortstack{Non-Default\\Loans}}
    & l.amt & 326.06 & 770.41 & 16.97 & 52.41 & 107.89 & 224.49 & 1497.81 \\
    & term & 1.60 & 1.50 & 1.00 & 1.00 & 1.00 & 1.00 & 6.00 \\
    & int.rate & 1.58 & 3.86 & 0.00 & 0.00 & 0.00 & 0.00 & 12.00 \\
    & l.itv & 18.58 & 32.82 & 0.00 & 0.00 & 5.00 & 21.00 & 87.00 \\
  \bottomrule
\end{tabular}
\end{table}

\begin{table}[t]
\centering
  \caption{Summary Statistics of Orders. The number of orders is 499,572, where 149,564 of them are in the sub-sequences for default loans. Variable names oamt, damt, qtty, catep, and oitv represent order amount (CNY), discount amount (CNY), quantity purchased, number of commodity category purchased, and time interval between consecutive orders (day), respectively.}
  \label{tab:su_order}
  \begin{tabular}{ccccccccc}
    \toprule
    & Variable & Mean & SD & 5th & 25th & Median & 75th & 95th \\
    \midrule
    \multirow{5}{*}{\shortstack{All\\Orders}}
    & oamt & 664.35 & 10746.74 & 28.53 & 57.06 & 171.18 & 399.42 & 3024.21 \\
    & damt & 77.55 & 288.74 & 0.00 & 0.00 & 0.00 & 57.06 & 313.83 \\
    & qtty & 2.73 & 14.52 & 1.00 & 1.00 & 1.00 & 2.00 & 7.00 \\
    & catep & 1.82 & 1.72 & 1.00 & 1.00 & 1.00 & 2.00 & 5.00 \\
    & oitv & 7.78 & 14.78 & 0.00 & 0.00 & 2.00 & 9.00 & 34.00 \\
    \midrule
    \multirow{5}{*}{\shortstack{Orders w.r.t.\\Default Loans}}
    & oamt & 579.33 & 7847.55 & 28.53 & 57.06 & 142.65 & 370.89 & 2995.68 \\
    & damt & 59.96 & 216.95 & 0.00 & 0.00 & 0.00 & 57.06 & 256.77 \\
    & qtty & 2.49 & 15.53 & 1.00 & 1.00 & 1.00 & 2.00 & 6.00 \\
    & catep & 1.69 & 1.61 & 1.00 & 1.00 & 1.00 & 2.00 & 5.00 \\
    & oitv & 6.34 & 13.88 & 0.00 & 0.00 & 1.00 & 6.00 & 30.00 \\
    \midrule
    \multirow{5}{*}{\shortstack{Orders w.r.t.\\Non-Default Loans}}
    & oamt & 700.68 & 11769.62 & 28.53 & 85.59 & 171.18 & 399.42 & 3195.39 \\
    & damt & 85.06 & 314.16 & 0.00 & 0.00 & 0.00 & 85.59 & 342.36 \\
    & qtty & 2.83 & 14.07 & 1.00 & 1.00 & 1.00 & 3.00 & 8.00 \\
    & catep & 1.87 & 1.76 & 1.00 & 1.00 & 1.00 & 2.00 & 5.00 \\
    & oitv & 8.40 & 15.11 & 0.00 & 0.00 & 2.00 & 10.00 & 36.00 \\
  \bottomrule
\end{tabular}
\end{table}

\begin{table}[t]
\centering
  \caption{Summary Statistics of Click Sessions. The number of sessions is 356,338, where 102,425 of them are in the sub-sequences for default loans. Variable names nclick, catev, duration, and sitv represent number of clicks, number of category visited, duration of the session (minute), and time interval between consecutive sessions (minute), respectively. Note that sitv has values that are less than 15 minutes. This is because the collection of clicks into sessions is done day by day, therefore a session with a sitv less than 15 minutes means that on the one hand the session before it happened yesterday near midnight, on the other hand, the current session happens today right after last midnight.}
  \label{tab:su_click}
  \begin{tabular}{ccccccccc}
    \toprule
    & Variable & Mean & SD & 5th & 25th & Median & 75th & 95th \\
    \midrule
    \multirow{4}{*}{\shortstack{All\\Sessions}}
    & nclick & 10.66 & 17.51 & 1.00 & 2.00 & 5.00 & 12.00 & 40.00 \\
    & catev & 1.94 & 1.55 & 1.00 & 1.00 & 1.00 & 2.00 & 5.00 \\
    & duration & 120.09 & 454.27 & 0.00 & 0.94 & 18.42 & 105.08 & 572.39 \\
    & sitv & 401.58 & 438.80 & 0.00 & 35.78 & 206.23 & 697.48 & 1312.65 \\
    \midrule
    \multirow{4}{*}{\shortstack{Sessions w.r.t.\\Default Loans}}
    & nclick & 11.61 & 20.11 & 1.00 & 2.00 & 5.00 & 13.00 & 44.00 \\
    & catev & 2.00 & 1.67 & 1.00 & 1.00 & 1.00 & 2.00 & 5.00 \\
    & duration & 122.47 & 436.29 & 0.00 & 0.94 & 19.84 & 109.09 & 582.31 \\
    & sitv & 396.00 & 436.61 & 0.00 & 35.92 & 199.27 & 682.12 & 1309.10 \\
    \midrule
    \multirow{4}{*}{\shortstack{Sessions w.r.t.\\Non-Default Loans}}
    & nclick & 10.28 & 16.33 & 1.00 & 2.00 & 5.00 & 12.00 & 38.00 \\
    & catev & 1.91 & 1.51 & 1.00 & 1.00 & 1.00 & 2.00 & 5.00 \\
    & duration & 119.13 & 461.33 & 0.00 & 0.71 & 17.71 & 103.19 & 568.00 \\
    & sitv & 403.82 & 439.66 & 0.00 & 35.67 & 208.83 & 703.43 & 1313.95 \\
  \bottomrule
\end{tabular}
\end{table}

Table \ref{tab:su_loan}, \ref{tab:su_order} and \ref{tab:su_click} present the descriptive statistics for some features of loans, orders, and click sessions. There are 38,182 loans for 5,000 consumers in our dataset, where 11,184 of the loans default. Each consumer has 7.64 loans on average. The major features of a loan include the loan amount, loan term, and interest rate. Besides, the time interval between the current loan and the last loan is also of interest. As the table shows, default loans tend to have smaller loan amounts, longer loan terms, higher interest rates, and shorter borrowing intervals. 

There are 499,572 orders for 38,182 loans in our dataset, where 149,564 of the orders are in the sub-sequences for default loans. Each loan has an order sub-sequence with 13.08 orders on average. The major features of an order include order amount, discount amount, the number of items purchased (Qtty.), the number of categories purchased (Cate. Purchase). Besides, the time interval between the current order and the last order is also of interest. As the table shows, default loans are usually related to orders with lower order amount, lower discount amount, fewer items and categories of products within an order, and shorter ordering intervals, suggesting the possibility of irrational consumption. 

There are 356,338 sessions for 38,182 loans in our dataset, where 102,425 of the sessions are in the sub-sequences for default loans. Each loan has a session sub-sequence with 9.33 sessions on average. The major features of a click session include the number of clicks within a session (N. of clicks), the number of categories visited (Cate. Visit), and the duration of a session. Besides, the time interval between the current session and the last session is also interesting. As the table shows, default loans are usually related to sessions with more clicks and more considerable duration of sessions, suggesting higher user stickiness.

\section{Methodology} \label{sec:method}
In this section, we introduce the NeuCredit model, which takes the temporal sequences of browsing, ordering, and borrowing as input and outputs the consumer credit risk at the issuance of each loan. The components of the model are illustrated one after another in the following subsections. We use bold lowercase letters to denote vectors and bold uppercase letters to denote matrices. A summary of variable notations is provided in Appendix \ref{app:notion}. The shapes of vectors and matrices can also be found in the summary.

\subsection{Input Definition}
For a consumer on an e-commerce platform, her borrowing actions forms a loan sequence $L=\{\bm{l}_i|i=1,2,...,T\}$ where $T$ is the time-stamp of loan issuance and $\bm{l}_i\in\mathbb{R}^{d_l}$ is the vector containing variables related to loan $i$. $d_l$ is the number of dimensions of $\bm{l}_i$. The loan variables are comprised of two parts: loan features such as amount, interest rate, loan term, etc., and a temporal feature specifying the time interval between this loan and the last loan. 

For each loan, ordering actions before and within a preset observation period are assigned to the loan to form a corresponding order sub-sequence. There are in total $T$ order sub-sequences $O=\{O_i|i=1,2,...,T\}$ where $O_i=\{\bm{o}_{i,1},\bm{o}_{i,2},...,\bm{o}_{i,|O_i|}\}$ is the order sub-sequence for loan $i$. $\bm{o}_{i,j}\in\mathbb{R}^{d_o}$ is the vector containing order information like order amount, product quantity, and the time interval between this order and the last order, etc. $d_o$ is the number of dimensions of $\bm{o}_{i,j}$. 

Browsing actions are first grouped into sessions, where a session is defined as beginning with a click which occurs after 15 minutes or more have elapsed since the last click and continuing until 15 minutes or more elapse between clicks. Then, the sessions are assigned to loans in the same manner as orders. This gives $T$ sub-sequences of browsing sessions $S=\{S_i|i=1,2,...,T\}$ where $S_i=\{\bm{s}_{i,1},\bm{s}_{i,2},...,\bm{s}_{i,|S_i|}\}$ is the browsing session sub-sequence for loan $i$. $\bm{s}_{i,j}\in\mathbb{R}^{d_s}$ is the vector containing the browsing information within session $j$ of loan $i$ such as duration of the session, time-on-page, total number of clicks, and the time interval between this session and the last one, etc. $d_s$ is the number of dimensions of $\bm{s}_{i,j}$. 

\begin{figure}[t]
\centering
\includegraphics[scale=0.7]{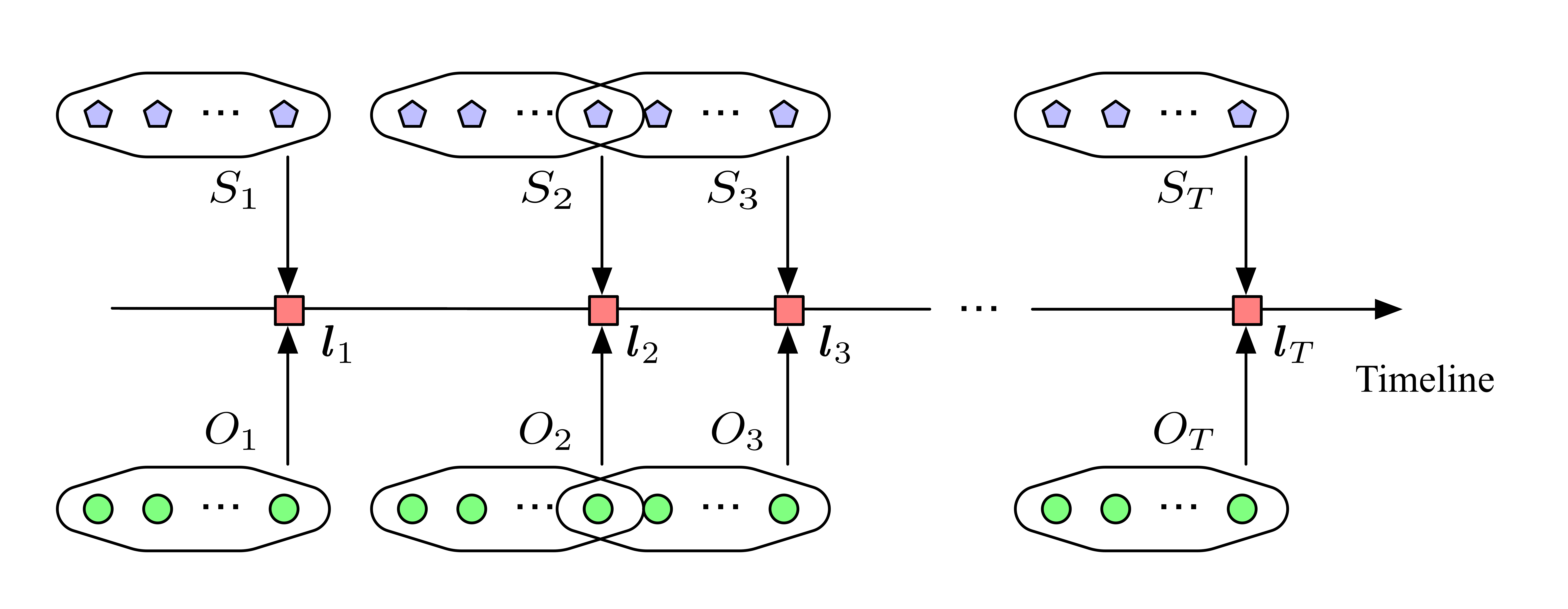}
\caption{An example of structured time-series consumer behavior. A series of loans is observed along the timeline for a consumer. Loans, orders, and sessions are marked red, green and purple, respectively. The time-stamps marked by red squares are at loan issuance. $\bm{l}_i$ is the vector containing loan features for loan $i$. Orders and sessions within preset observation periods before loan issuance are grouped and assigned to loans to form two sub-sequences for each loan. $O_i$ is a set of order vectors that forms the order sub-sequence for loan $i$. $S_i$ is a set of session vectors that forms the session sub-sequence for loan $i$. Sub-sequences will overlap with each other if loans cluster in time (as sub-sequences for loan 2 and 3 in the figure). Time intervals between consecutive elements in each sequence are not equal with each other due to behavioral irregularities in time.}
\label{fig:action}
\end{figure}

An exemplary data structure is illustrated in Figure \ref{fig:action}.

\subsection{Sequence Encoding}
The most fundamental component of NeuCredit is the recurrent unit employed to learn behavioral dynamics. Usually, Long Short-Term Memory (LSTM) neural network \citep{hochreiter1997long,gers1999learning} is regarded as the most popular and effective recurrent unit in plenty of sequence modeling tasks \citep{ren2015exploring,wang2016attention,yang2017attention}. However, conventional sequential models, including LSTM, implicitly assume that elements in a sequence are discrete and uniformly distributed along the timeline, i.e., time intervals between consecutive elements are equal. This is not the case in most real-life tasks where events happen stochastically in continuous time. Time intervals between consumer actions can reveal valuable information in many scenarios, including credit risk modeling. For instance, a recent purchase of an expensive good in cash indicates a good economic condition, while a purchase months ago may not play an active role in predicting the default risk of the current loan issued to finance an order.

In our situation, events in a loan sequence $L$ as well as in its related order sub-sequences $O$ and session sub-sequences $S$ are taking place irregularly in time. So it is imperative to consider these irregularities in modeling. In the literature, the most straightforward approach is to regard the time interval between two successive elements in a sequence as an extra feature so that the standard LSTM is applicable as before. As Equation (\ref{eq:lstm}) shows, this approach implicitly models the non-linear effects of the time interval on other features through the activation functions in LSTM. 

In Equation (\ref{eq:lstm}), $\odot$ is the Hadamard product operator that implements the element-wise multiplication, $\sigma(\cdot)$ and $\operatorname{tanh}(\cdot)$ are activation functions that introduce non-linearity into fitting, $\bm{x}_t$ represents the current input vector, $\Delta t$ is the time interval between the current time-stamp and the previous time-stamp, $\bm{h}_{t-1}$ and $\bm{h}_t$ are the previous and the current hidden states, $\bm{c}_{t-1}$ and $\bm{c}_t$ are the previous and the current cell memories, \{$\bm{W}_i$, $\bm{U}_i$, $\bm{b}_i$\}, \{$\bm{W}_f$, $\bm{U}_f$, $\bm{b}_f$\}, \{$\bm{W}_o$, $\bm{U}_o$, $\bm{b}_o$\}, and \{$\bm{W}_c$, $\bm{U}_c$, $\bm{b}_c$\} are the trainable network parameters of the input, forget, output gates and the candidate memory, respectively, and $\bm{i}_t$, $\bm{f}_t$, $\bm{o}_t$, and $\tilde{\bm{c}}_t$ are the input, forget, output gates and the candidate memory, respectively. 

\begin{equation}
\begin{aligned}
\label{eq:lstm}
    & \bm{i}_t = \sigma(\bm{W}_i [\bm{x}_t;\Delta t] + \bm{U}_i \bm{h}_{t-1} + \bm{b}_i) \\
    & \bm{f}_t = \sigma(\bm{W}_f [\bm{x}_t;\Delta t] + \bm{U}_f \bm{h}_{t-1} + \bm{b}_f) \\
    & \bm{o}_t = \sigma(\bm{W}_o [\bm{x}_t;\Delta t] + \bm{U}_o \bm{h}_{t-1} + \bm{b}_o) \\
    & \tilde{\bm{c}}_t = \operatorname{tanh}(\bm{W}_c [\bm{x}_t;\Delta t] + \bm{U}_c \bm{h}_{t-1} + \bm{b}_c) \\
    & \bm{c}_t = \bm{f}_t \odot \bm{c}_{t-1} + \bm{i}_t \odot \tilde{\bm{c}}_t \\
    & \bm{h}_t = \bm{o}_t \odot \operatorname{tanh}(\bm{c}_t)
\end{aligned}
\end{equation}
The shape of these vectors and matrices are in Appendix \ref{app:notion}. For the theories and details of Long Short-Term Memory neural network, please refer to \citet{hochreiter1997long} and \citet{gers1999learning}.

Alternatively, \citet{baytas2017patient} is the first to explicitly model the effect of time intervals by proposing Time-aware LSTM (T-LSTM). Instead of regarding $\Delta t$ as a common feature, the authors use it to process the cell memory $\bm{c}_{t-1}$ in standard LSTM. Specifically, the cell memory $\bm{c}_{t-1}$ is first decomposed into short-term and long-term memories. Then, the short-term memory is discounted by a factor $g(\Delta t)$ where $g(\cdot)$ is some preset monotonically non-increasing function. The long-term and the discounted short-term memories are next fused into $\bm{c}'_{t-1}$ that serves the role of the original cell memory in standard LSTM. The mathematical forms of the above operations are as follows,
\begin{equation}
\begin{aligned}
\label{eq:kdd}
    & \bm{c}_{t-1}^S = \operatorname{tanh}(\bm{W}_D \bm{c}_{t-1}+\bm{b}_D) \\
    & \bm{c}_{t-1}^L = \bm{c}_{t-1} - \bm{c}_{t-1}^S \\
    & \bm{c}_{t-1}^{S'} = \bm{c}_{t-1}^S * g(\Delta t) \\
    & \bm{c}'_{t-1} = \bm{c}_{t-1}^L + \bm{c}_{t-1}^{S'}
\end{aligned}
\end{equation}

In Equation \eqref{eq:kdd}, $\bm{c}_{t-1}$ is the cell memory in standard LSTM, $\bm{c}_{t-1}^S$ and $\bm{c}_{t-1}^L$ are the short-term and long-term memories, respectively, $\bm{c}_{t-1}^{S'}$ is the discounted short-term memory, $\bm{W}_D$ and $\bm{b}_D$ are trainable network parameters for decomposition, and $\bm{c}'_{t-1}$ is the new cell memory that will take the place of the original $\bm{c}_{t-1}$ in Equation (\ref{eq:lstm}). According to \citet{baytas2017patient}, T-LSTM performs much better than standard LSTM on both synthetic and real world sequential data.

However, this method is problematic to some extent. First, it uses a preset function $g(\cdot)$ that only allows monotonically non-increasing discounting of the cell memory and thus prohibits the enhancement of cell memory in time. This setting is too rigorous in practice as some events are effective in a very long run, and their importance can even naturally grows over time. 

For instance, the amount of money deposited in a bank can increase persistently at the interest rate. Second, the third formula in Equation (\ref{eq:kdd}) implicitly assumes that the values at different positions of vector $\bm{c}_{t-1}^S$ possess a same discounting rate $g(\Delta t)$, which limits the expressiveness of T-LSTM. Third, the discounting is taking place in a low-dimensional space which makes it hard for $g(\Delta t)$ to discount information in high dimensions. This constraint is caused by the network parameter $\bm{W}_D$ that maintains the number of dimensions during mapping. Lastly, the discounting with a preset function $g(\cdot)$ lacks theoretical insights about how does $\Delta t$ come into effect in modeling.

Therefore, we propose Time-value-aware LSTM (Tva-LSTM) that settles the problems of T-LSTM. Tva-LSTM is very flexible that allows both decaying and growing of the cell memory over time. The decaying or growing rates are trainable so that the discounting process is data-driven. The discounting is taking place in a high-dimensional space, and each dimension has its own discounting rate. 

Besides, the discounting mechanism is theoretically derived upon a reasonable assumption so that it shades lights on the functionality of $\Delta t$. Particularly, the cell memory vector $\bm{c}_{t-1}$ is first mapped to a high-dimensional space represented by a matrix $\bm{C}_{t-1}$. At the same time, a discounting matrix $\bm{D}_{t-1}$ that has the same shape as $\bm{C}_{t-1}$ is initialized by $\Delta t$. Then, $\bm{D}_{t-1}$ multiplies $\bm{C}_{t-1}$ element-wisely to allow different discounting rates for different dimensions. 

Lastly, the product matrix $\bm{C}_{t-1}^D$ is mapped back to a low-dimensional space to serve as the new cell memory $\bm{c}'_{t-1}$. The non-linearity is introduced via activation functions. The mathematical forms of the above operations are as follows,
\begin{equation}
\begin{aligned}
\label{eq:tva}
    & \bm{C}_{t-1} = \operatorname{tanh}(\bm{c}_{t-1} \bm{w}_H + \bm{B}_H) \\
    & \bm{D}_{t-1} = e^{\operatorname{tanh}(\bm{W}_R * \Delta t + \bm{B}_R)} \\
    & \bm{C}_{t-1}^D = \operatorname{tanh}(\bm{C}_{t-1} \odot \bm{D}_{t-1} + \bm{B}_D) \\
    & \bm{c}'_{t-1} = \operatorname{tanh}(\bm{C}_{t-1}^D \bm{w}_L + \bm{b}_L)
\end{aligned}
\end{equation}

In Equation \eqref{eq:tva}, $\bm{c}_{t-1}$ is the cell memory in standard LSTM, $\bm{C}_{t-1}$ is the mapped cell memory in a high-dimensional space, $\bm{D}_{t-1}$ is the corresponding discounting matrix, $\bm{C}_{t-1}^D$ is the discounted mapped cell memory, and $\bm{c}'_{t-1}$ is the new cell memory that will take the place of the original $\bm{c}_{t-1}$ in Equation (\ref{eq:lstm}). \{$\bm{w}_H$, $\bm{B}_H$\} are the trainable parameters responsible for mapping the cell memory to a high-dimensional space. \{$\bm{W}_R$, $\bm{B}_R$\} are the trainable parameters for initializing the discounting matrix. $\bm{B}_D$ is the trainable parameter for discounting the mapped cell memory. \{$\bm{w}_L$, $\bm{b}_L$\} are the trainable parameters for mapping the discounted mapped cell memory back to a low-dimensional space. 

Note that the discounting factor $\bm{D}_{t-1}$ takes the form of exponentiation. In fact, this specific form can be derived by assuming that the elements in the mapped cell memory are continuously changing at different rates over time. Since the derivation is straightforward, we put it in Appendix \ref{app:discounting} for clarity.

\begin{figure}[t]
	\centering
	\includegraphics[scale=0.5]{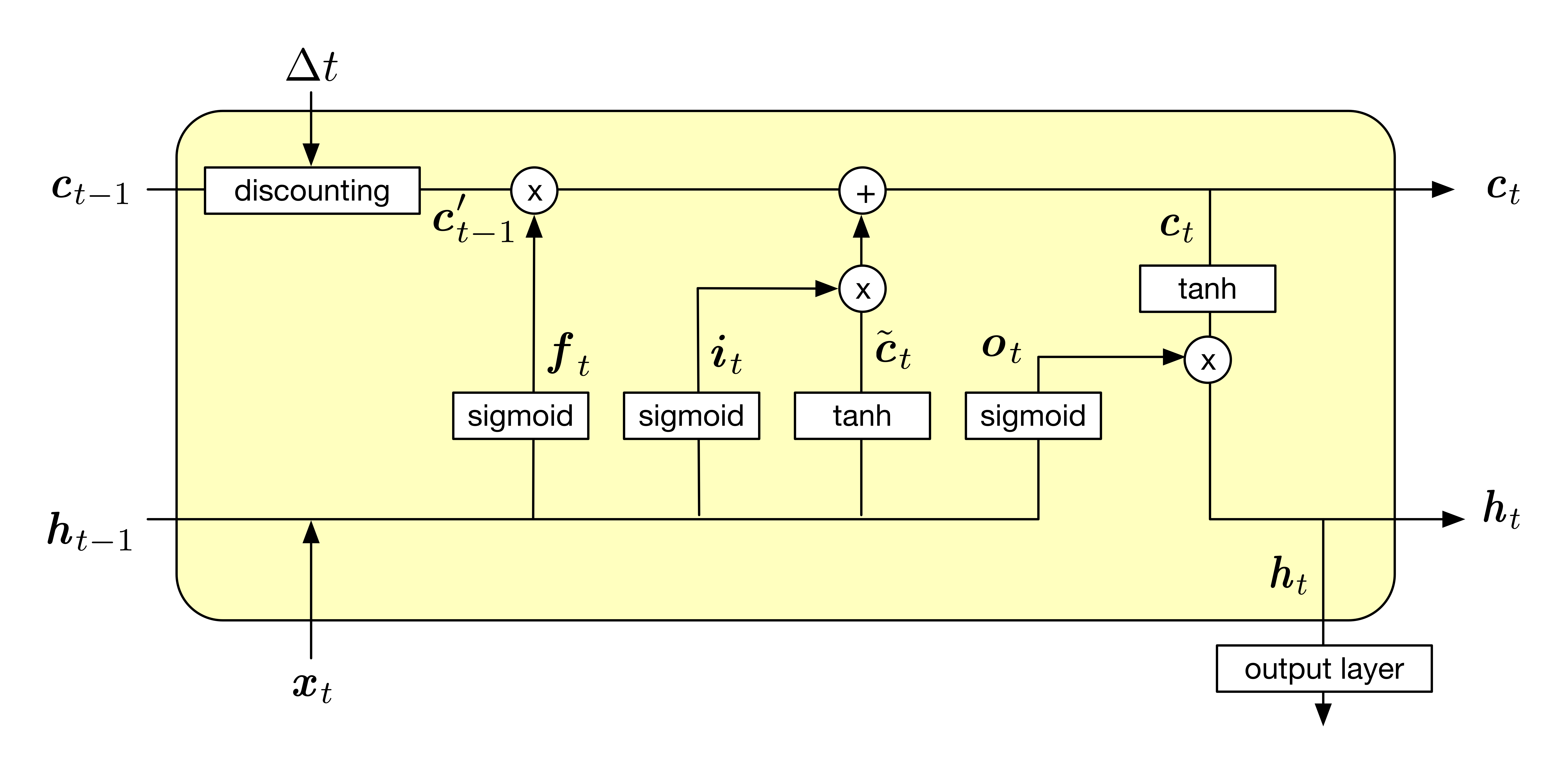}
	\caption{The illustration of the Tva-LSTM recurrent unit. The product sign with a circle around denotes the point-wise multiplication operator and the plus sign with a circle around is the point-wise addition operator. The operations inside the discounting module are presented in Equation (\ref{eq:tva}). Other operations are the same as Equation (\ref{eq:lstm}). The sigmoid and the tanh represent the activation function $\sigma(\cdot)$ and $\operatorname{tanh}(\cdot)$, respectively.}
	\label{fig:tva_lstm}
\end{figure}

Figure \ref{fig:tva_lstm} gives a brief illustration of the proposed Tva-LSTM recurrent unit. To be specific, Tva-LSTM takes the hidden state $\bm{h}_{t-1}$ and the cell memory $\bm{c}_{t-1}$ from last moment as inputs. Before passing them to different gates, the cell memory first entries into a discounting unit to regularize the time gap between the last moment and the current moment. In the discounting unit, the cell memory $\bm{c}_{t-1}$ will first be mapped into a high-dimensional space, then be element-wisely discounted via a discounting factor matrix, and lastly be mapped back to the original low-dimensional space. 

As denoted in Equation (\ref{eq:tva}), the complete process of time gap regularization is data-driven such that both the mapping parameters and the decaying/growing rate parameters are learned simultaneously with the rest of network parameters by back-propagation. This renders Tva-LSTM very expressive as it not only allows both decaying and growing of cell memory over time but also assigns different changing rates to different dimensions in the high-dimensional space. Following discounting, the hidden state $\bm{h}_{t-1}$ and the regularized cell memory $\bm{c}'_{t-1}$ are passed to typical LSTM gates.

\subsection{Multi-view Fusion}
Another critical component of the NeuCredit model is the fusion strategy used to combine the main loan sequence and its related sub-sequences. The objective of fusion is to integrate the information heterogeneity maintained in different views of actions and more importantly, to model the mutual effects due to behavioral interactions. In this study, order and session sub-sequences are encoded via two Tva-LSTM, separately. The fusion is carried out at the issuance of each loan in the loan sequence.

Taking the fusion at loan $i$ as an example, the inputs of fusion are the loan vector $\bm{l}_i$, the final hidden state $\bm{h}^{o}_{i,|O_i|}$ of the Tva-LSTM for the $i$-th order sub-sequence $O_i$, and the final hidden state $\bm{h}^{s}_{i,|S_i|}$ of the Tva-LSTM for the $i$-th session sub-sequence $S_i$. One straightforward idea is to first concatenate the three vectors and then pass it through a fully connected neural network layer with a nonlinear activation function $\sigma(\cdot)$, i.e.,
\begin{equation}
\label{eq:fc}
    \bm{z}_i=\sigma(\bm{W}_F[\bm{l}_i;\bm{h}^{o}_{i,|O_i|};\bm{h}^{s}_{i,|S_i|}]+\bm{b}_F).
\end{equation}

Another approach is to capture the interactions of different groups of actions by exploiting the concept of Multi-view Machines \citep{cao2016mvm}. Here, we employ a Multi-view Machines layer \citep{cao2017deepmood} for fusion. The layer explicitly models the feature interactions so that it acquires non-linearity more efficiently in training. Besides, it captures full-order interactions from 0 to the number of input vectors. For the theories and details of Multi-view Machines, please refer to \citet{cao2016mvm,cao2017deepmood}. The formula of this layer is
\begin{equation}
\label{eq:mvm}
    \bm{z}_i = (\bm{U}_{F1} [\bm{l}_i;1]) \odot (\bm{U}_{F2} [\bm{h}^{o}_{i,|O_i|};1]) \odot (\bm{U}_{F3} [\bm{h}^{s}_{i,|S_i|};1]),
\end{equation}
where $\bm{U}_{F1}$, $\bm{U}_{F2}$, and $\bm{U}_{F3}$ are three trainable factor matrices for fusion. Their shapes are $(d_z,d_l+1)$, $(d_z,d_{ho}+1)$, and $(d_z,d_{hs}+1)$, respectively. $d_{ho}$ and $d_{hs}$ are the number of hidden units in the Tva-LSTM for order sub-sequences and session sub-sequence, respectively. $d_z$ is the number of dimensions of the fused vector $\bm{z}_i$.

\subsection{Hierarchical Network}
\begin{figure}[t]
\centering
\includegraphics[scale=0.55]{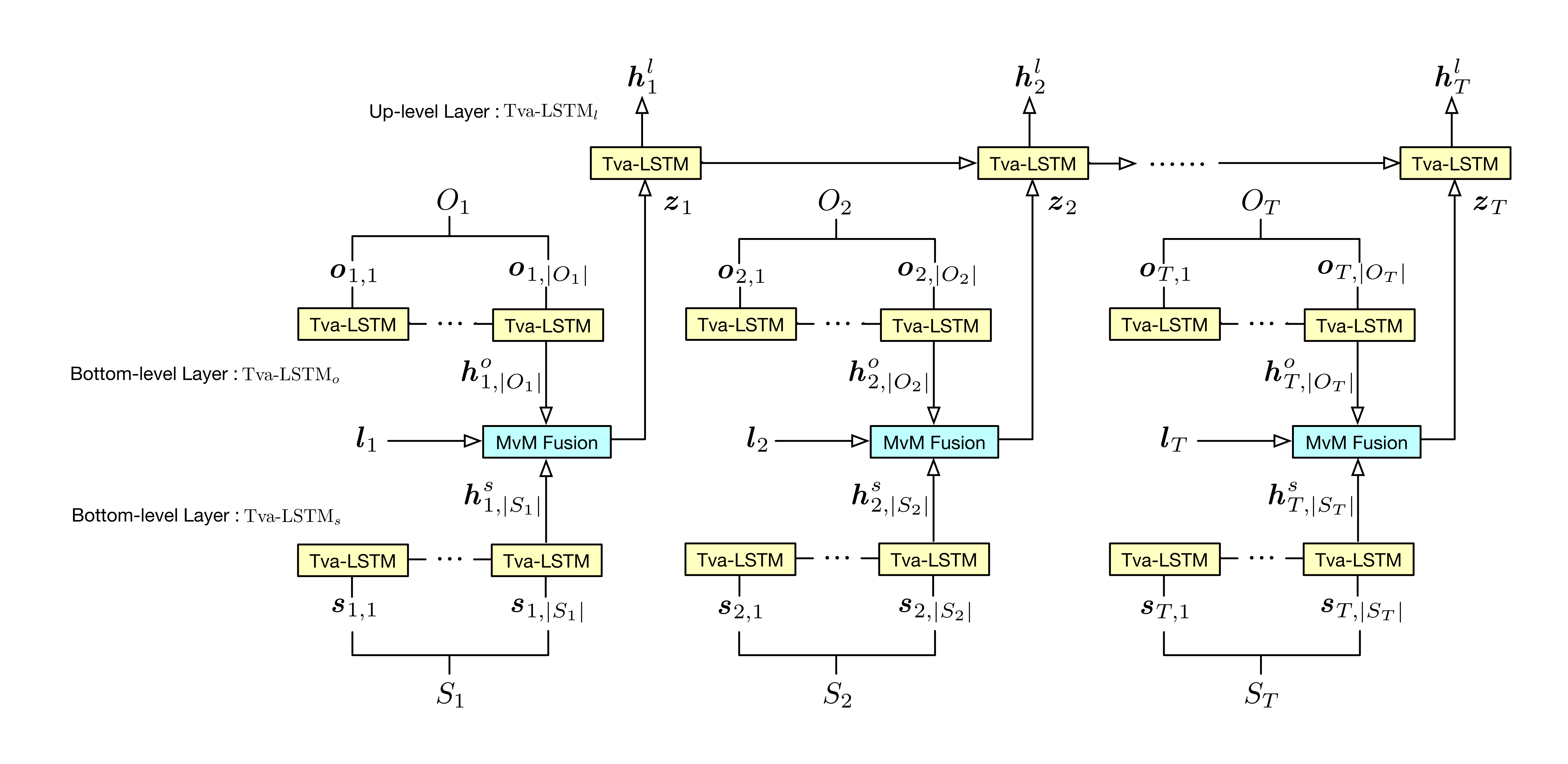}
\caption{The architecture of the hierarchical network.}
\label{fig:neucredit}
\end{figure}

In this part, the forementioned components are combined to present the hierarchical network proposed for sophisticated consumer behavior. The architecture is illustrated in Figure \ref{fig:neucredit}. 

In the bottom-level layers, two separate Tva-LSTM recurrent units are used to encode the order sub-sequences and session sub-sequences. This avoids the difficulties of aligning different groups of actions that have distinct patterns of serial-dependency and frequencies of occurrence. For sub-sequences $O_i$ and $S_i$, the encoding is done as follows,
\begin{equation}
\begin{aligned}
\label{eq:tva_lstm_for_bottom_level}
    & \bm{h}^o_{i,j}=\operatorname{Tva-LSTM}_o(\bm{h}^o_{i,j-1},\bm{o}_{i,j}),\forall\bm{o}_{i,j}\in O_i \\
    & \bm{h}^s_{i,j}=\operatorname{Tva-LSTM}_s(\bm{h}^s_{i,j-1},\bm{s}_{i,j}),\forall\bm{s}_{i,j}\in S_i,
\end{aligned}
\end{equation}
where $\bm{h}^o_{i,j}\in\mathbb{R}^{d_{ho}}$ and $\bm{h}^s_{i,j}\in\mathbb{R}^{d_{hs}}$ are the hidden states of Tva-LSTM units, $\operatorname{Tva-LSTM}_o$ and $\operatorname{Tva-LSTM}_s$ denote the two Tva-LSTM units employed for order and session sub-sequences. The last hidden states $\bm{h}^{o}_{i,|O_i|}$ and $\bm{h}^{s}_{i,|S_i|}$ summarize the information in sub-sequences $O_i$ and $S_i$ and thus are regarded as their final representations.

In the up-level layer, the loan vector $\bm{l}_i$ is first fused with $\bm{h}^{o}_{i,|O_i|}$ and $\bm{h}^{s}_{i,|S_i|}$ as in Equation (\ref{eq:mvm}). The procedure is denoted by the \textit{MvM Fusion} unit in Figure \ref{fig:neucredit}. Following that, the fused vector $\bm{z}_i$ is encoded by a up-level Tva-LSTM:
\begin{equation}
\label{eq:tva_lstm_for_up_level}
    \bm{h}^l_i=\operatorname{Tva-LSTM}_l(\bm{h}^l_{i-1},\bm{z}_i),\forall i\in \{1,2,...,T\},
\end{equation}
where $\bm{h}^l_i\in\mathbb{R}^{d_{hl}}$ is the hidden state of the up-level recurrent unit $\operatorname{Tva-LSTM}_l$. $d_{hl}$ is the number of hidden units. $\bm{h}^l_i$ represents a summary of consumer behavior up to time-stamp $i$ in the loan sequence $L$.

\subsection{Conditional Loss}
In the last section, we successfully obtain the representation $\bm{h}^l_i$ of all historical events at time-stamp $i$. Following that, $\bm{h}^l_i$ is often used to fulfill some classification or regression tasks. For example, in credit management, a critical task for risk assessment is to predict whether a loan will default. A loan is considered as default if its repayment delays more than 90 days. The prediction can be implemented as follows,
\begin{equation}
\label{eq:pre_dp}
    \hat{P}_d\coloneqq\hat{y}_i=\sigma(\bm{w}_P\bm{h}^l_i+b_P),
\end{equation}

In Equation \eqref{eq:pre_dp}, $\bm{w}_P$ is a trainable vector that maps $\bm{h}^l_i$ to one dimension, $b_P$ is a real-value bias, $\sigma(\cdot)$ is the sigmoid activation function, and the predicted default probability $\hat{P}_d$ is $\hat{y}_i$. The dissimilarity between $\hat{y}_i$ and the real binary outcome $y_i$ is measured by a loss function $\ell_1(\hat{y}_i,y_i)$. $y_i=1$ if loan $i$ defaults; otherwise, $y_i=0$. The model parameters are learned by minimizing the loss function in training.

This approach is standard in classification problems. But it has one serious drawback in credit risk modeling. The predicted default probability is not interpretable. It neither distinguishes the sources of risk nor illuminates the contributions of different sources to default. Here, we propose to construct the default probability based on three major determinants of loan defaults \citep{lee1991ability,chehrazi2015dynamic}: the objective risk (the ability to repay), the subjective risk (the willingness to repay), and the behavioral risk (the risk neither objective nor subjective). In probability, we formulate the default probability as follows,
\begin{equation}
\label{eq:dp_decomp}
    P_d\coloneqq P(b|a,w)P(a)P(w),
\end{equation}

In Equation \eqref{eq:dp_decomp}, $P_d$ is the default probability, $P(a)$ is the default probability when the ability is $a$, $P(w)$ is the default probability when the willingness is $w$, and $P(b|a,w)$ is the default probability conditioned on $a$ and $w$, i.e., the default risk caused by behavioral patterns other than the ability and the willingness to repay. In this way, the default probability becomes interpretable.

To simulate the construction of an interpretable default probability in neural networks, we first decompose $\bm{h}^l_i$ into three vectors:
\begin{equation}
\begin{aligned}
\label{eq:a_w_split}
    & \bm{h}^a_i = \operatorname{tanh}(\bm{W}_A \bm{h}^l_i+\bm{b}_A) \\
    & \bm{h}^w_i = \operatorname{tanh}(\bm{W}_W \bm{h}^l_i+\bm{b}_W) \\
    & \bm{h}^b_i = \bm{h}^l_i - \bm{h}^a_i - \bm{h}^w_i,
\end{aligned}
\end{equation}
where $\{\bm{W}_A,\bm{W}_W,\bm{b}_A,\bm{b}_W\}$ are trainable parameters for decomposition, and $\bm{h}^a_i$, $\bm{h}^w_i$, and $\bm{h}^b_i$ are hidden vectors containing the information for ability risk, willingness risk, and behavioral risk, respectively. Then, the hidden vectors are separately mapped to one dimension to predict $P(a)$, $P(w)$, and $P(b|a,w)$:
\begin{equation}
\begin{aligned}
\label{eq:map_to_a_w_b}
    & \hat{P}(a)\coloneqq\hat{y}_i^a = \sigma(\bm{w}_A \bm{h}^a_i+b_A) \\
    & \hat{P}(w)\coloneqq\hat{y}_i^w = \sigma(\bm{w}_W \bm{h}^w_i+b_W) \\
    & \hat{P}(b|a,w)\coloneqq\hat{y}_i^b = \sigma(\bm{w}_B\bm{h}^b_i+b_B),
\end{aligned}
\end{equation}
where $\{\bm{w}_A,\bm{w}_W,\bm{w}_B,b_A,b_W,b_B\}$ are trainable parameters for mapping. Following that, the predicted default probability $\hat{y}_i=\hat{y}_i^a\hat{y}_i^w\hat{y}_i^b$ is supervised by $\ell_1(\hat{y}_i,y_i)$ as before. 

In order to let $\hat{y}_i^a$, $\hat{y}_i^w$, and $\hat{y}_i^b$ truly represent the meaning we imposed on them, it is imperative to supervise them independently by their own ground-truth in training. However, $P(a)$, $P(w)$, and $P(b|a,w)$ are completely unobservable in practice. Therefore, we put forward a method to infer the values of $P(a)$ and $P(w)$ for a loan by carefully analyzing the repayment behavior on that loan. 

Particularly, if a borrower defaults on a loan, although we are not sure about whether it is caused by a low ability or a low willingness to repay, we can still infer that one of them must be low enough to lead to the outcome. That is, in probability, the probability of the default that is caused by neither ability nor willingness to repay is very low. 

On the contrary, if a borrower repays every installment on time and never defaults on that loan, it is certain that he has not only a high ability but also a high willingness to repay. Another interesting situation in between is that a borrower never defaults, but he is often delinquent (overdue) on the periodical installments of that loan. In this condition, the repaying ability of the borrower must be high as he is always able to complete the payment, but the willingness may be low because of her frequent delinquencies. Mathematically, the inference above can be summarized as
\begin{equation}
\label{eq:inference}
\begin{cases}
    (1-\hat{P}(a))(1-\hat{P}(w))=0,& \text{if } y_i=1 \\
    \hat{P}(a)=0,\hat{P}(w)=0,     & \text{if } y_i=0,r_i=0 \\
    \hat{P}(a)=0,\hat{P}(w)=r_i,   & \text{if } y_i=0,r_i>0
\end{cases}
\end{equation}
where $r_i\in[0,1]$ is the proportion of the installments of loan $i$ that the borrower has been delinquent on. In this way, we inferred the values of $P(a)$ and $P(w)$ under different conditions. These inferred values can be used as teachers in training via a conditional loss function:
\begin{equation}
\label{eq:con_loss}
\ell_2=
\begin{cases}
    (1-\hat{y}_i^a)(1-\hat{y}_i^w),& \text{if } y_i=1 \\
    (\hat{y}_i^a)^2+ (r_i-\hat{y}_i^w)^2,& \text{if } y_i=0
\end{cases}
\end{equation}

Note that $\ell_2$ is conditioned on a binary variable $y_i$, we can write the two expressions into one and combine it with $\ell_1$. In summary, the proposed loss function for the NeuCredit model is
\begin{equation}
\label{eq:full_loss}
\ell=\sum_{1}^b\sum_{i=1}^T\{\ell_1(\hat{y}_i,y_i)+y_i(1-\hat{y}_i^a)(1-\hat{y}_i^w)+(1-y_i)[(\hat{y}_i^a)^2+(r_i-\hat{y}_i^w)^2]\}
\end{equation}
where $b$ is the batch size used in mini-batch optimization and $T$ is the loan sequence length. The first part of Equation (\ref{eq:full_loss}) is the conventional loss for classification. Here, we use binary cross-entropy as $\ell_1(\cdot)$. The second and third parts of Equation (\ref{eq:full_loss}) are the conditional loss hinging on the value of $y_i$.

Following the computational graph, one can straightforwardly compute the gradients for all the network parameters in the NeuCredit model. Also, the error messages by weighing the predicted outputs with the observed loan outcomes can be back-propagated through the decomposition layers and fusion layers all the way to the very beginning to update the parameters in different branches of Tva-LSTMs. In that sense, the NeuCredit model is said to be an end-to-end deep neural network model that learns the dynamics of consumer behavior for interpretable credit risk modeling.

\section{Experiment} \label{sec:experiment}
In this section, we design and conduct experiments using both synthetic datasets and real-life datasets to address the following four groups of questions:
\begin{itemize}
	\item How much better are deep learning models than conventional machine learning models?
	\item How much value is added by incorporating shopping behavior data when forecasting consumer credit risk?
	\item Is it indeed important to model the irregular event time-internals?
	\item Can we interpret the forecasted default probabilities into consumer's ability to repay, willingness to repay, and their behavioral factors?
\end{itemize}

We use the synthetic dataset to demonstrate the superiority of the Tva-LSTM model over other competing ones on recovering the dynamics of complex patterns. The construction details is in Appendix \ref{app:synthetic}. The synthetic dataset contains 10,000 sequences with a length of 50 for each sequence. Every data point in the dataset has 106 features and 1 label. Among the 106 features, only 5 are involved in the generation of the label, while the rest is all noise. Besides, to produce sequential dependencies, the 5 features at the current time-stamp in a sequence is generated by transforming the 5 features at the previous time-stamp in a highly non-linear manner. The label of each data point is a binary indicator which takes the value of 1 or 0. Among the 500,000 data points, 323,326 of them are positive instances, i.e., their labels equal to 1.   

The real-life dataset contains 5,000 loan sequences with 38,182 loans in total. The average length of loan sequences is 7.64. Each loan possesses 15 features ($d_l=15$). Among the 38,182 loans, 11,184 ($29.29\%$) of them default. For each loan, an order sub-sequence and a session sub-sequence are matched. Therefore, there are 5,000 order sub-sequences and 5,000 session sub-sequences. The dataset contains 499,572 orders and 356,338 sessions. On average, the length of an order sub-sequence is 13.08 and of a session sub-sequence is 9.33. Each order possesses 45 features ($d_o=45$) and each session possesses 16 features ($d_s=16$).

In the experiments, sequences and sub-sequences with length less than 15 are padded to length of 15 using 0. The influence of padding is eliminated through masking both in training and testing. This treatment is a common practice in temporal data modeling, which allows us to handle variable length sequences in recurrent models. Features are standardized before passing to models. 

As different group of questions require a different set of benchmark models, these models and their implementation details are left to be specified in corresponding subsections. All methods are evaluated using five-fold cross-validation \citep{kohavi1995study}. The Area-Under-ROC Curve (AUC) score is used as the primary performance metric in evaluation \citep{bradley1997use}. Experiments are implemented using Python. Pandas\footnote{http://pandas.pydata.org/} and Numpy\footnote{http://www.numpy.org/} libraries are used to process the datasets. Scikit-Learn\footnote{http://scikit-learn.org/} and Tensorflow\footnote{https://www.tensorflow.org/} libraries are used to implement the algorithms. The source code of all implementation will be publicly available after paper acceptance.

\subsection{Deep Learning vs. Conventional Machine Learning Models}
In this part, we test the performance improvements of our model over other conventional and competitive models. Specifically, does our model perform better in credit risk prediction than conventional models? Can a model with a similar structure but conventional units achieve comparable performance to our model? To answer these questions, the following methods are compared in experiments:

\begin{itemize}
	\item \textbf{LR (loan)}: the Logistic Regression model trained on loans with the time interval as an extra feature. This is similar to the traditional consumer credit management scenario where only financing behavior can be observed.
	\item \textbf{LR (all)}: the Logistic Regression model trained on all three groups of data (loans, orders, and sessions). The features of sub-sequences are averaged along the timeline and concatenated with loan features. The time intervals are regarded as extra features.
	\item \textbf{RF (loan)}: the Random Forest model trained on loans with the time interval as an extra feature. This is similar to the traditional consumer credit management scenario where only financing behavior can be observed.
	\item \textbf{RF (all)}: the Random Forest model trained on all three groups of data (loans, orders, and sessions). The features of sub-sequences are averaged along the timeline and concatenated with loan features. The time intervals are regarded as extra features.
	\item \textbf{LSTM-w-dt (loan)}: the standard LSTM model trained on loans with the time interval as an extra feature.
	\item \textbf{MvM-Tva-LSTM (all)}: the model that employs the same hierarchical structure and fusion mechanism as Figure \ref{fig:neucredit}. The model is trained on all three groups of sequential data (loans, orders, and sessions).
\end{itemize}

The models are trained to predict loan defaults using binary cross-entropy loss. The number of hidden units ($d_h$) is set as 5 for the Tva-LSTM unit and the LSTM unit employed in the aforementioned methods. The number of output units ($d_z$) of the fully-connected fusion layer in the FC-LSTM model is set as 5. The number of output units ($d_z$) of the factor matrices in the MvM-Tva-LSTM model is set as 5. All neural network models are trained with a mini-batch stochastic Adam optimizer \citep{kingma2014adam}. The batch size is set as 1,000. The learning rate is 0.001. The number of epochs in training is determined using an early stopping criteria \citep{caruana2001overfitting}. The logistic regression models and the random forest models are trained with default parameter setting in Scikit-Learn. The AUCs of different models in five-fold cross-validation are shown in Table \ref{tab:performance_pk}. 

First, the conventional methods indeed cannot reach comparable performance to deep neural network methods. Second, compared with the FC-LSTM model that employs conventional units but uses the same hierarchical structure as that of our model, the MvM-Tva-LSTM model achieves better performance in experiments. An interesting finding is that the average AUC of the FC-LSTM is $73.43\%$, which outperforms the average AUC of the FC-Tva-LSTM model in Section \ref{subsec:interact}. This is inconsistent with our finding in Section \ref{subsec:action_irregularity} that the Tva-LSTM model is better at handling the time intervals and can outperform the conventional LSTM model without $\Delta t$. The reason is that both the FC-LSTM model and the FC-Tva-LSTM model are trained in an end-to-end manner that requires a model to learn all the parameters from scratch (cold-start). While the units in the FC-LSTM model are conventional and easy to train, the units in the FC-Tva-LSTM model are much more complicated in design. It leads to insufficient training of the Tva-LSTM unit in the FC-Tva-LSTM model. This problem can be settled by using the pre-trained parameters to initialize the Tva-LSTM in FC-Tva-LSTM (warm-start). In general, these results demonstrate the feasibility and effectiveness of using shopping behavior to model credit risk for consumers. Also, it suggests that the proposed hierarchical architecture is better at capturing the underlying behavioral patterns of consumers than conventional methods.

\begin{table}[t]
	\centering
	\caption{AUCs Achieved with Different Models in Five-fold Cross-validation}
	\label{tab:performance_pk}
	\begin{tabular}{cccccc|cc}
		\toprule
		Method/AUC (\%) & AUC-1 & AUC-2 & AUC-3 & AUC-4 & AUC-5 & Avg. AUC & S.D. \\
		\midrule
		LR (loan) & 63.59 & 64.68 & 66.33 & 60.47 & 63.96 & 63.80 & 0.0191 \\
		LR (all) & 69.13 & 68.99 & 71.22 & 67.02 & 68.18 & 68.91 & 0.0138 \\
		RF (loan) & 59.76 & 61.72 & 60.17 & 59.83 & 60.80 & 60.46 & 0.0073 \\
		RF (all) & 68.59 & 67.16 & 69.28 & 66.97 & 68.53 & 68.11 & 0.0089 \\
		\midrule
		LSTM-w-dt (loan) & 70.59 & 70.45 & 69.87 & 68.30 & 71.69 & 70.18 & 0.0111 \\
		MvM-Tva-LSTM (all) & \textbf{74.25} & \textbf{73.22} & \textbf{75.86} & \textbf{72.37} & \textbf{73.98} & \textbf{73.94} & 0.0116 \\
		\bottomrule
	\end{tabular}
\end{table}

\subsection{The Importance of Adding Browsing and Purchasing Data}
\label{subsec:interact}
To better understand the roles played by different views of shopping behavior in default risk modeling, we train a Tva-LSTM model on each of the three types of temporal data. Besides, we study the importance of modeling the behavioral interactions and the necessity of multi-view fusion. Specifically, without borrowing data, are consumer behavior alone contain information in terms of predicting the outcomes of borrowing? If they are, does the multi-view fusion strategy successfully model the behavioral interactions in online shopping and uncover their contributions to credit risk prediction? Does the Multi-view Machines fusion layer behave better than the straightforward fully-connected fusion layer? To answer these questions, the following methods are implemented in experiments:

\begin{itemize}
    \item \textbf{Tva-LSTM (loan)}: the Time-value-aware LSTM model trained on loan sequences.
    \item \textbf{Tva-LSTM (order)}: the Time-value-aware LSTM model trained on order sub-sequences, the hierarchical structure as Figure \ref{fig:neucredit} is employed without fusion with other sequences/sub-sequences.
    \item \textbf{Tva-LSTM (session)}: the Time-value-aware LSTM model trained on session sub-sequences, the hierarchical structure as Figure \ref{fig:neucredit} is employed without fusion with other sequences/sub-sequences.
    \item \textbf{FC-Tva-LSTM (all)}: the model that employs the same hierarchical structure as Figure \ref{fig:neucredit} but uses a fully-connected layer instead of a Multi-view Machines layer for fusion. The model is trained on all three groups of sequential data (loans, orders, and sessions).
    \item \textbf{MvM-Tva-LSTM (all)}: the model that employs the same hierarchical structure and fusion mechanism as Figure \ref{fig:neucredit}. The model is trained on all three groups of sequential data (loans, orders, and sessions).
\end{itemize}

\begin{table}[t]
\centering
  \caption{AUCs Achieved with Different Real-Life Data Streams in Five-fold Cross-validation}
  \label{tab:sv_vs_mv}
  \begin{tabular}{cccccc|cc}
   \toprule
    Method/AUC (\%) & AUC-1 & AUC-2 & AUC-3 & AUC-4 & AUC-5 & Avg. AUC & S.D. \\
    \midrule
    Tva-LSTM (loan) & 71.13 & 71.14 & 71.02 & 68.95 & 72.29 & 70.91 & 0.0108 \\
    Tva-LSTM (order) & 72.53 & 71.28 & 72.29 & 69.87 & 72.42 & 71.68 & 0.0101 \\
    Tva-LSTM (click) & 54.88 & 53.87 & 54.61 & 57.27 & 54.43 & 55.01 & 0.0118 \\
    \midrule
    FC-Tva-LSTM (all) & 73.11 & 72.04 & 74.87 & 71.18 & 73.92 & 73.02 & 0.0131 \\
    MvM-Tva-LSTM (all) & \textbf{74.25} & \textbf{73.22} & \textbf{75.86} & \textbf{72.37} & \textbf{73.98} & \textbf{73.94} & 0.0116 \\
  \bottomrule
\end{tabular}
\end{table}

The models are trained to predict loan defaults using binary cross-entropy loss. The number of hidden units ($d_h$) is set as 5 for all Tva-LSTM units employed in the aforementioned methods. The number of output units ($d_z$) of the fully-connected fusion layer in the FC-Tva-LSTM model is set as 5. The number of output units ($d_z$) of the factor matrices in the MvM-Tva-LSTM model is set as 5. All models are trained with a mini-batch stochastic Adam optimizer \citep{kingma2014adam}. The batch size is set as 1,000. The learning rate is 0.001. The number of epochs in training is determined using an early stopping criteria \citep{caruana2001overfitting}. The AUCs of different models in five-fold cross-validation are shown in Table \ref{tab:sv_vs_mv}. 

First, the average AUCs achieved with orders and sessions are $71.68\%$ and $55.01\%$. Both of them are greater than $50\%$, indicating that behavior in shopping is indeed useful in predicting one's credit risk. Moreover, ordering actions seem more informative than borrowing actions as the average AUC achieved with orders is consistently higher than that achieved with loans. This finding provides the empirical foundation that supports the development of online credit shopping on e-commerce platforms. When all types of actions are considered, the performance increases. The AUCs of both the FC-Tva-LSTM model and the MvM-Tva-LSTM model are consistently higher than that of models with a single type of actions, indicating that the fusion layer indeed explored the macroscopic interactions cross different views of data, and the mutual effects it exposed are effective in evaluating consumer credit risk. Besides, the MvM-Tva-LSTM model performs better than the FC-Tva-LSTM model, which demonstrates Multi-view Machines fusion is better at capturing the interactions.

\subsection{The Importance of Modeling Irregular Time-intervals of Events}
\label{subsec:action_irregularity}
In this part, we study the importance of handling irregularity in temporal data modeling. Specifically, is it indeed necessary to take time intervals into consideration? Is the proposed handling of time intervals, via the Tva-LSTM model, better at capturing the irregularities in behavior than other methods? To answer these questions, the following models are used for comparison:

\begin{itemize}
	\item \textbf{LSTM}: the standard LSTM model that ignores time intervals.
	\item \textbf{LSTM-w-dt}: the standard LSTM model that takes time intervals into modeling as in (\ref{eq:lstm}).
	\item \textbf{T-LSTM}: the Time-aware LSTM model proposed by \citet{baytas2017patient} that takes time intervals into modeling via a present discounting function $g(\Delta t)$.
	\item \textbf{Tva-LSTM}: the Time-value-aware LSTM model proposed in this study that handles the time intervals in a more expressive way.
\end{itemize}

\begin{figure}[t]
	\centering
	\includegraphics[scale=0.38]{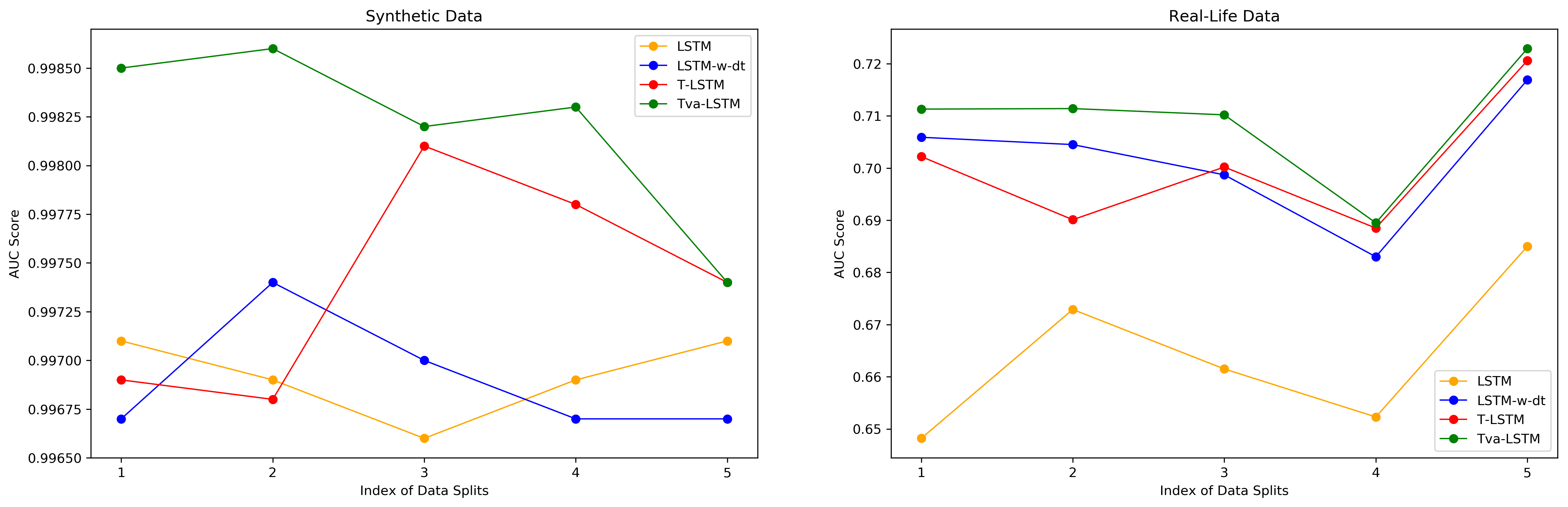}
	\caption{AUCs of Sequential Models in Five-fold Cross-validation}
	\label{fig:lstm_comp}
\end{figure}

The experiments are conducted on both the synthetic sequences and the loan sequences in the real-life data. The models are trained to predict loan defaults using binary cross-entropy loss. The number of hidden units ($d_h$) is set as 2 for all models in experiments with the synthetic data, and 5 for all models in experiments with the real-life data. All models are trained with a mini-batch stochastic RMSprop optimizer \citep{mukkamala2017variants}. The batch size is set as 1,000. The learning rate is 0.001. The number of epochs in training is determined using an early stopping criteria \citep{caruana2001overfitting}. The AUCs of different models in five-fold cross-validation are plotted in Figure \ref{fig:lstm_comp}. The average AUCs are shown in Table \ref{tab:irregularity}. 

The proposed Tva-LSTM model achieves the best performance on both synthetic and real-life data. Models that incorporate time intervals achieve better average AUC in experiments. This implication is more evident in experiments with real-life data, where LSTM-w-dt, T-LSTM, and Tva-LSTM all outperform the conventional LSTM by more than three percentage points. These results demonstrate the necessity of taking the time intervals into consideration and the superiority of the proposed discounting mechanism in Tva-LSTM.

\begin{table}[t]
	\centering
	\caption{Prediction Performance of Sequential Models}
	\label{tab:irregularity}
	\begin{tabular}{ccccc}
		\toprule
		Data & \multicolumn{2}{c}{Synthetic} & \multicolumn{2}{c}{Real-Life} \\
		\midrule
		Method/Metric & Avg. AUC (99\%+bps) & S.D. & Avg. AUC (\%) & S.D. \\
		\midrule
		LSTM & 69 & 0.0002 & 66.40 & 0.0135 \\
		LSTM-w-dt & 69 & 0.0003 & 70.18 & 0.0111 \\
		T-LSTM & 74 & 0.0005 & 70.03 & 0.0115 \\
		Tva-LSTM & \textbf{82} & 0.0004 & \textbf{70.91} & 0.0108 \\
		\bottomrule
	\end{tabular}
\end{table}

\subsection{Structural Interpretation of Forecasted Default Probabilities}
Up to now, models such as FC-LSTM, FC-Tva-LSTM, or MvM-Tva-LSTM are all trained to predict loan defaults using binary cross-entropy loss. In this part, we turn to the interpretable conditional loss function and evaluate the complete NeuCredit model. Specifically, we want to address the following questions: Given the highly complicated structures and operations inside the NeuCredit model, does it converge properly in training? If it does, what is its performance? More importantly, are the values of predicted ability and predicted willingness consistent with our design? How does consumer behavior relate to the ability and the willingness of repaying?

\begin{figure}[t]
	\centering
	\includegraphics[scale=0.38]{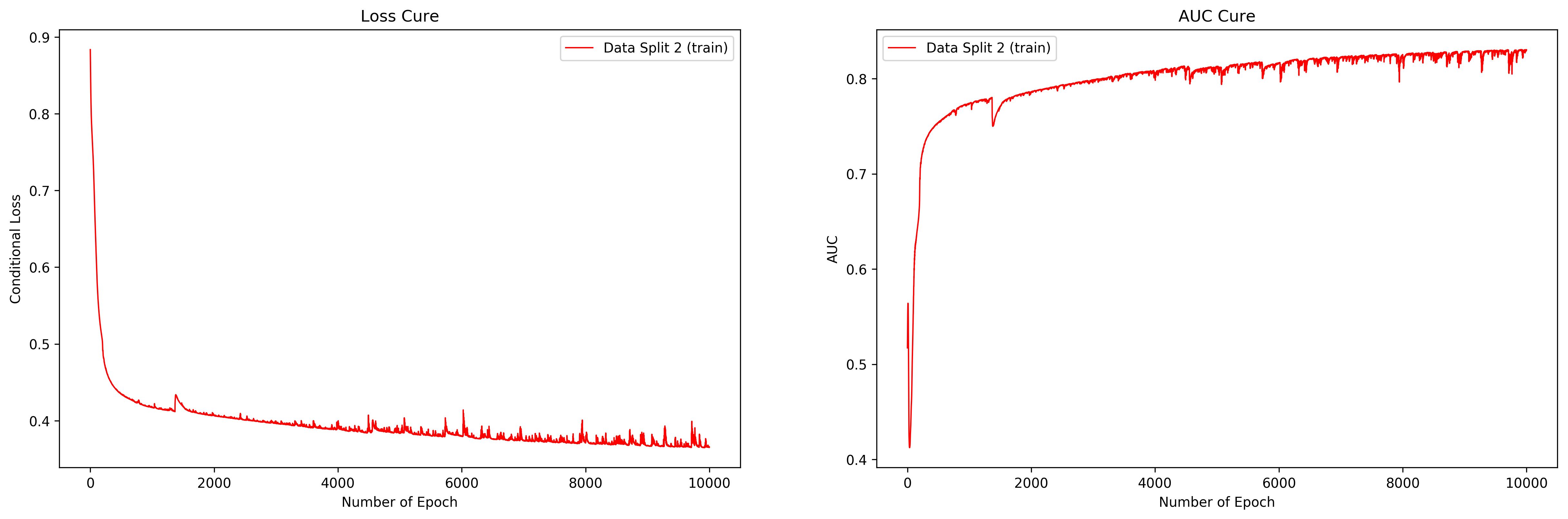}
	\caption{The Loss Curves and the AUC Curves in Training and Validation}
	\label{fig:loss_auc}
\end{figure}

The parameter setting of the NeuCredit model is the same as that of the MvM-Tva-LSTM model. The curve of training loss and the curve of training AUC are plotted in Figure \ref{fig:loss_auc}. Here,  only the curves for one of the five splits in five-fold cross-validation is presented for simplicity. As they show, the convergence of the NeuCredit model is not affected even if we employ many complicated units such as Tva-LSTM, Multi-view Machines Fusion, and Conditional Loss in the NeuCredit model. The loss is continuously decreasing and the AUC is continuously increasing as the training process proceeds. 

Besides, the prediction performance is presented in Table \ref{tab:neucredit}. The performance of the MvM-Tva-LSTM model is also presented in Table \ref{tab:neucredit} for reference. Note that the performance of the NeuCredit model is inferior to MvM-Tva-LSTM. It is a reasonable result since while MvM-Tva-LSTM is trained to serve as an MLE (maximum likelihood estimator), i.e., directly optimizing the binary cross-entropy, the NeuCredit model needs to weigh between the prediction performance and the interpretability. This trade-off leads to a little performance decrease of the NeuCredit model in default risk prediction. 

\begin{table}[t]
	\centering
	\caption{AUCs Achieved with the NeuCredit model in Five-fold Cross-validation}
	\label{tab:neucredit}
	\begin{tabular}{cccccc|cc}
		\toprule
		Method/AUC (\%) & AUC-1 & AUC-2 & AUC-3 & AUC-4 & AUC-5 & Avg. AUC & S.D. \\
		\midrule
		NeuCredit & 74.91 & 72.18 & 72.39 & 74.00 & 75.06 & 73.71 & 0.0122 \\
		MvM-Tva-LSTM & 74.25 & 73.22 & 75.86 & 72.37 & 73.98 & 73.94 & 0.0116 \\
		\bottomrule
	\end{tabular}
\end{table}

\begin{figure}[h]
	\centering
	\includegraphics[scale=0.51]{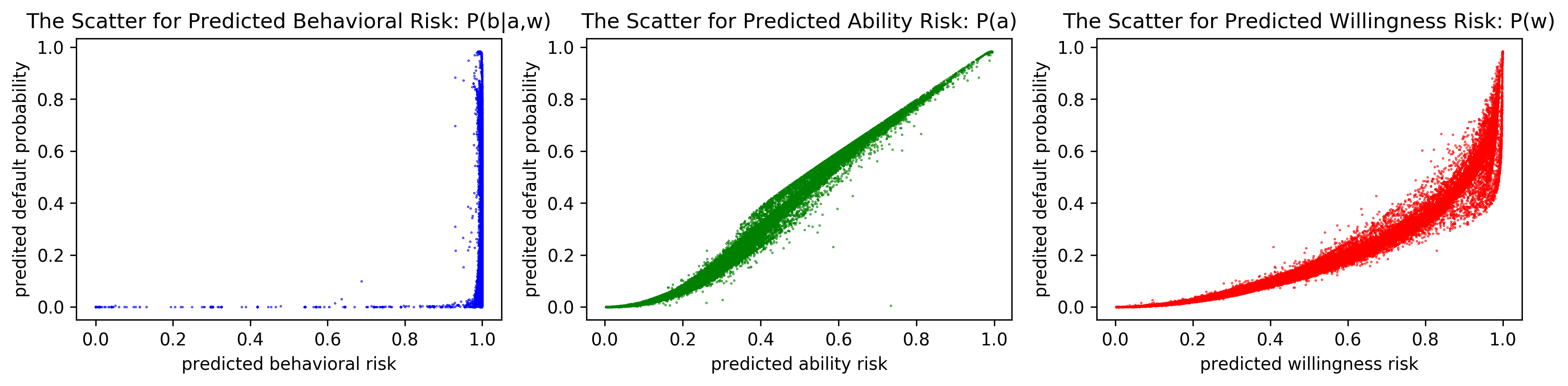}
	\caption{The Scatter for Predicted Default Determinants}
	\label{fig:scatter}
\end{figure}

Next, we check if the predicted values of behavioral risk, ability risk, and willingness risk are consistent with our design. We interpret results from two perspectives. First, the three types of predicted risks are scattered against predicted default probabilities on Figure \ref{fig:scatter} to visualize the correlation between credit risk and its determinants. Together with equation \eqref{eq:dp_decomp}, it is very interesting to notice how the ultimate default probability attributes to the three claimed risk types. 

We then use linear regression to test if the three types of risk are significantly correlated with the outcomes of loans, i.e., the default indicator $y$ and the delinquency ratio $r$. The meaning of all regression variables is detailed in Table \ref{tab:variable}. The results are collected in Table \ref{tab:reg_part_1}, \ref{tab:reg_part_2}, and \ref{tab:reg_part_3}. The regressions in Table \ref{tab:reg_part_1} reveals what we have encoded in the NeuCredit model. The regressions in Table \ref{tab:reg_part_2} answers the question of whether the predicted values are significant in differentiating consumers with different risk levels. Finally, the regressions in Table \ref{tab:reg_part_3} studies if the obtained factors are significant determinants in predicting consumer defaults. 

As it shows, the behavioral risk and the willingness risk are indeed positively correlated with $y$ and $r$. The predicted values of the willingness risk is consistent with our expectation. One thing worthy of future investigations is that the correlation between the ability risk and default or delinquency ratio is reversed. Also, the explanatory power of the predicted ability risk is low compared to the other two types of predicted risk. The reason could be that the model does not distinguish well the behavioral risk from the ability risk as they essentially share the same guidance $y$ in the training process. And the reason that the predicted behavioral risk is in line with our expectation is that it is supervised by $\ell_1(\hat{y}^b\hat{y}^a\hat{y}^w,y)$ which incorporates more information via $\hat{y}^w$.  In general, both the scatter plots and the regressions demonstrate that the predicted values of three types of determinants are consistent with our design and they do reveal the sources of credit risk.

\begin{landscape}
\begin{table}[h]
	\centering
	\caption{Explanation of Regression Variables}
	\label{tab:variable}
	\begin{tabular}{ccc}
		\toprule
		Group & Variable & Description \\
		\midrule
		\multirow{1}{*}{Real Outcome}
		& $y$ & the dummy variable that indicates default of a loan \\
		\midrule
		\multirow{7}{*}{\shortstack{Prediction of\\NeuCredit}}
		& $\hat{y}$ & the predicted default probability of a loan \\
		& $\hat{y}^a$ & the predicted default probability of a loan with ability $=a$ \\
		& $\hat{y}^w$ & the predicted default probability of a loan with willingness $=w$ \\
		& $\hat{y}^b$ & the predicted default probability of a loan with behavioral risk $=b$ \\
		& $dummy^a$ & 1 if $\hat{y}^a$ is larger than the median of $\hat{y}^a$; 0 otherwise \\
		& $dummy^w$ & 1 if $\hat{y}^w$ is larger than the median of $\hat{y}^w$; 0 otherwise \\
		& $dummy^b$ & 1 if $\hat{y}^b$ is larger than the median of $\hat{y}^b$; 0 otherwise \\
		\midrule
		\multirow{7}{*}{\shortstack{Loan\\Variable}}
		& $\log lamt$ & the natural logarithm of the principal of a loan (CNY) \\
		& $term$ & the term of a loan (month) \\
		& $intrate$ & the annualized interest rate of a loan \\
		& $\Delta t^l$ & the time interval between the current and the previous loan issuance \\
		& $\log mnpay$ & the natural logarithm of the minimum payment of a loan installment (CNY) \\
		\midrule
		\multirow{9}{*}{\shortstack{Order\\Variable}}
		& $\log oamt$ & the natural logarithm of the average order amount (CNY) \\
		& $disrate$ & the average discount rate of an order \\
		& $qtty$ & the average quantity of items within an order \\
		& $item$ & the average quantity of different items within an order \\
		& $\Delta t^o$ & the average time interval between the current and the previous order \\
		& $vgp$ & the average proportion of virtual goods within an order  \\
		& $sgp$ & the average proportion of self-selling goods within an order \\
		& $fgp$ & the average proportion of free gifts within an order \\
		& $lv$ & the user level when placing an order \\
		\midrule
		\multirow{4}{*}{\shortstack{Session\\Variable}}
		& $click$ & the average number of clicks within a session \\
		& $catev$ & the average number of category visited \\
		& $duration$ & the average duration of a session (minute) \\
		& $\Delta t^s$ & the average time interval between the current and the previous session \\
		\bottomrule
	\end{tabular}
\end{table}
\end{landscape}

\newpage
\begin{table}[h]
	\centering
	\caption{Regression: What we've encoded via the NeuCredit model?}
	\label{tab:reg_part_1}
	\begin{tabular}{clllll}
		\toprule
		\multicolumn{1}{c}{Explanatory/Response} & \multicolumn{1}{c}{$y$} & \multicolumn{1}{c}{$\hat{y}$} & \multicolumn{1}{c}{$\hat{y}^a$} & \multicolumn{1}{c}{$\hat{y}^w$} & \multicolumn{1}{c}{$\hat{y}^b$} \\
		\midrule
		$\log lamt$ & $0.7096^{***}$ & $0.0064^{***}$ & $0.0064^{***}$ & $0.0048^{***}$ & $-0.0002$\\
		$term$ & $0.2637^{***}$ & $0.0193^{***}$ & $0.0164^{***}$ & $0.0094^{***}$ & $-0.0002^{*}$\\
		$intrate$ & $0.8879$ & $0.3487^{***}$ & $0.3244^{***}$ & $0.5396^{***}$ & $0.0129^{*}$ \\
		$\Delta t^l$ & $0.0014^{**}$ & $-0.0004^{***}$ & $-0.0004^{***}$ & $-0.0004^{***}$ & $0.0000$\\
		$\log mnpay$ & $-0.3908^{***}$ & $-0.0051^{***}$ & $-0.0047^{***}$ & $-0.0025^{*}$ & $0.0002^{*}$\\
		$\log oamt$ & $0.0655$ & $0.0065^{***}$ & $0.0050^{***}$ & $0.0083^{***}$ & $0.0007^{*}$ \\
		$disrate$ & $-0.1945$ & $0.0064$ & $0.0061$ & $0.0078$ & $-0.0018$ \\
		$qtty$ & $0.0022$ & $-0.0000$ & $-0.0000$ & $0.0000$ & $0.0002$\\
		$item$ & $-0.0612$ & $-0.0050^{**}$ & $-0.0035^{**}$ & $-0.0067^{**}$ & $0.0007$\\
		$\Delta t^o$ & $0.0376^{***}$ & $0.0024^{***}$ & $0.0019^{***}$ & $0.0034^{***}$ & $0.0000$ \\
		$vgp$ & $-0.3209$ & $-0.0645^{***}$ & $-0.0509^{***}$ & $-0.0687^{***}$ & $-0.0016$ \\
		$sgp$ & $0.2851^{*}$ & $0.0195^{***}$ & $0.0200^{***}$ & $0.0315^{***}$ & $-0.0019$ \\
		$fgp$ & $-0.1370$ & $0.0178^{*}$ & $0.0102$ & $0.0122$ & $-0.0014$ \\
		$lv$ & $-1.1958^{***}$ & $-0.0316^{***}$ & $-0.0141^{***}$ & $-0.0384^{***}$ & $0.0008$\\
		$click$ & $-0.0002$ & $-0.0001$ & $-0.0001$ & $-0.0000$ & $-0.0000$ \\
		$catev$ & $-0.0503$ & $0.0027^{*}$ & $0.0018$ & $0.0034^{*}$ & $0.0006^{*}$ \\
		$duration$ & $0.0007$ & $-0.0002$ & $-0.0002$ & $-0.0000$ & $0.0000$ \\
		$\Delta t^s$ & $-0.0027$ & $-0.0003$ & $-0.0002$ & $-0.0006^{*}$ & $0.0000$\\
		\midrule
		\# of obs. & $15,399$ & $38,160$ & $38,160$ & $38,160$ & $38,160$ \\
		\# of groups & $1,762$ & $4,998$ & $4,998$ & $4,998$ & $4,998$ \\
		\bottomrule
	\end{tabular}
\end{table}

\newpage
\begin{table}[h]
	\centering
	\caption{Regression: Are the predicted values significant in differentiating consumers with different risk levels?}
	\label{tab:reg_part_2}
	\begin{tabular}{clllll}
		\toprule
		\multicolumn{1}{c}{Explanatory/Response} & \multicolumn{1}{c}{$y$} & \multicolumn{1}{c}{$y$} & \multicolumn{1}{c}{$y$} & \multicolumn{1}{c}{$y$} & \multicolumn{1}{c}{$y$}\\
		\midrule
		$dummy^b$ & & $0.1200$ \\
		$dummy^a$ & & & $0.1545^{**}$ & & $-0.3228^{**}$ \\
		$dummy^w$ & & & & $0.2651^{***}$ & $0.1302$ \\
		$dummy^a*dummy^w$ & & & & & $0.4295^{**}$ \\
		\midrule
		$\log lamt$ & $0.7096^{***}$ & $0.7096^{***}$ & $0.7074^{***}$ & $0.7066^{***}$ & $0.7059^{***}$ \\
		$term$ & $0.2637^{***}$ & $0.2651^{***}$ & $0.2627^{***}$ & $0.2626^{***}$ & $0.2623^{***}$ \\
		$intrate$ & $0.8879$ & $0.8198$ & $0.7245$ & $0.6221$ & $0.6636$ \\
		$\Delta t^l$ & $0.0014^{**}$ & $0.0012^{*}$ & $0.0016^{**}$ & $0.0016^{**}$ & $0.0015^{**}$ \\
		$\log mnpay$ & $-0.3908^{***}$ & $-0.3906^{***}$ & $-0.3899^{***}$ & $-0.3893^{***}$ & $-0.3888^{***}$ \\
		$\log oamt$ & $0.0655$ & $0.0658$ & $0.0614$ & $0.0607$ & $0.0643$ \\
		$disrate$ & $-0.1945$ & $-0.1894$ & $-0.2076$ & $-0.2055$ & $-0.1907$ \\
		$qtty$ & $0.0022$ & $0.0020$ & $0.0020$ & $0.0018$ & $0.0019$ \\
		$item$ & $-0.0612$ & $-0.0604$ & $-0.0560$ & $-0.0502$ & $-0.0484$ \\
		$\Delta t^o$ & $0.0376^{***}$ & $0.0376^{***}$ & $0.0367^{***}$ & $0.0360^{***}$ & $0.0360^{***}$ \\
		$vgp$ & $-0.3209$ & $-0.3190$ & $-0.3154$ & $-0.3026$ & $-0.2992$ \\
		$sgp$ & $0.2851^{*}$ & $0.2833^{*}$ & $0.2734^{*}$ & $0.2699^{*}$ & $0.2693^{*}$ \\
		$fgp$ & $-0.1370$ & $-0.1411$ & $-0.1320$ & $-0.1413$ & $-0.1471$ \\
		$lv$ & $-1.1958^{***}$ & $-1.1921^{***}$ & $-1.1893^{***}$ & $-1.1845^{***}$ & $-1.1767^{***}$ \\
		$click$ & $-0.0002$ & $-0.0002$ & $-0.0002$ & $-0.0001$ & $-0.0001$ \\
		$catev$ & $-0.0503$ & $-0.0507$ & $-0.0513$ & $-0.0533$ & $-0.0540$ \\
		$duration$ & $0.0007$ & $0.0006$ & $0.0007$ & $0.0007$ & $0.0006$ \\
		$\Delta t^s$ & $-0.0027$ & $-0.0028$ & $-0.0023$ & $-0.0021$ & $-0.0024$ \\
		\midrule
		\# of obs. & $15,399$ & $15,399$ & $15,399$ & $15,399$ & $15,399$ \\
		\# of groups & $1,762$ & $1,762$ & $1,762$ & $1,762$ & $1,762$ \\
		\bottomrule
	\end{tabular}
\end{table}

\newpage
\begin{table}[h]
	\centering
	\caption{Regression: Are the obtained factors significant determinants in predicting consumer defaults?}
	\label{tab:reg_part_3}
	\begin{tabular}{clllll}
		\toprule
		\multicolumn{1}{c}{Explanatory/Response} & \multicolumn{1}{c}{$y$} & \multicolumn{1}{c}{$y$} & \multicolumn{1}{c}{$y$} & \multicolumn{1}{c}{$y$} & \multicolumn{1}{c}{$y$}\\
		\midrule
		$\hat{y}^b_{t-1}$ & & $1.1644$ & & & $0.5292$ \\
		$\hat{y}^a_{t-1}$ & & & $0.4886^{**}$ & & $-2.0444^{***}$ \\
		$\hat{y}^w_{t-1}$ & & & & $0.9028^{***}$ & $2.1665^{***}$ \\
		\midrule
		$\log lamt$ & $0.7096^{***}$ & $0.8042^{***}$ & $0.8058^{***}$ & $0.8093^{***}$ & $0.8086^{***}$ \\
		$term$ & $0.2637^{***}$ & $0.8042^{***}$ & $0.2978^{***}$ & $0.2985^{***}$ & $0.3018^{***}$ \\
		$intrate$ & $0.8879$ & $0.4057$ & $0.4021$ & $0.3682$ & $0.3478$ \\
		$\Delta t^l$ & $0.0014^{**}$ & $0.0001$ & $-0.0000$ & $-0.0004$ & $-0.0004$ \\
		$\log mnpay$ & $-0.3908^{***}$ & $-0.4957^{***}$ & $-0.4963^{***}$ & $-0.4983^{***}$ & $-0.4983^{***}$ \\
		$\log oamt$ & $0.0655$ & $0.0828^{*}$ & $0.0783$ & $0.0701$ & $0.0713$ \\
		$disrate$ & $-0.1945$ & $-0.1531$ & $-0.1634$ & $-0.1788$ & $-0.1753$ \\
		$qtty$ & $0.0022$ & $-0.0004$ & $-0.0001$ & $-0.0001$ & $-0.0003$ \\
		$item$ & $-0.0612$ & $-0.0109$ & $-0.0032$ & $0.0143$ & $0.0170$ \\
		$\Delta t^o$ & $0.0376^{***}$ & $0.0430^{***}$ & $0.0415^{***}$ & $0.0387^{***}$ & $0.0390^{***}$ \\
		$vgp$ & $-0.3209$ & $-0.1653$ & $-0.1618$ & $-0.1533$ & $-0.1515$ \\
		$sgp$ & $0.2851^{*}$ & $0.3344^{*}$ & $0.3164^{*}$ & $0.2774$ & $0.2814$ \\
		$fgp$ & $-0.1370$ & $-0.4277^{*}$ & $-0.4249^{*}$ & $-0.4398^{*}$ & $-0.4658^{*}$ \\
		$lv$ & $-1.1958^{***}$ & $-1.1232^{***}$ & $-1.113^{***}$ & $-1.0728^{***}$ & $-1.0418^{***}$ \\
		$click$ & $-0.0002$ & $-0.0000$ & $0.0000$ & $0.0001$ & $-0.0001$ \\
		$catev$ & $-0.0503$ & $-0.0493$ & $-0.0471$ & $-0.0495$ & $-0.0551$ \\
		$duration$ & $0.0007$ & $0.0027$ & $0.0031$ & $0.0034$ & $0.0031$ \\
		$\Delta t^s$ & $-0.0027$ & $-0.0042$ & $-0.0039$ & $-0.0031$ & $-0.0026$ \\
		\midrule
		\# of obs. & $15,399$ & $12,341$ & $12,341$ & $12,341$ & $12,341$ \\
		\# of groups & $1,762$ & $1,499$ & $1,499$ & $1,499$ & $1,499$ \\
		\bottomrule
	\end{tabular}
\end{table}

\newpage
\section{Conclusion} \label{sec:conclusion}
In this paper, we take a data-driven bottom-up approach to model consumer credit risk with structural interpretability in the e-commerce scenario when a platform provides unsecured lending to finance consumer purchasing and needs to manage the resulting credit exposure. By zooming into the tick-level shopping behavior and the subsequent financing records of large population, we open a window to profile consumer credit at an unprecedented granular level. Deciphering them carefully would allow real-time assessment of future payment risk, particularly when payments are financed without posting collateral. 

The structure of our deep neural network is novel. First, we propose Tva-LSTM recurrent unit to encode temporal shopping behavior that happen stochastically in time. Tva-LSTM unit effectively regularizes the time intervals in temporal data. The discounting mechanism in this unit is explainable as it is derived on mild assumptions. Then, the encoded representations are passed to a Multi-view Machines layer to do information fusion. The fusion strategy explicitly computes the interactions across different types of shopping behavior via tensor multiplication. Finally, the NeuCredit model organizes temporal data in a hierarchical structure which avoids dominant view problem and achieves real-time fusion of various types of information. Besides, we propose a novel conditional loss function that exploits repaying behavior to infer the values of determinants for credit risk. We decompose the consumer credit risk into three of its determinants: behavioral risk, ability-to-repay risk, and willingness-to-repay risk. The supervising of these risks are accomplished in training even if their ground-truths are not observable. In this way, the NeuCredit model is able to output interpretable credit risk predictions. Extensive experiments are conducted using both a synthetic dataset and a massive real-life dataset collected from one of the largest global e-commerce platforms. The out-of-sample forecasts of consumer default risk demonstrate the effectiveness of the methodology proposed in this paper, in terms of the superiority of our model over conventional machine learning models as well as other state-of-the-art deep learning models, as well as the interpretability of the model predictions.

In our opinion, there are three future directions that are very interesting to study further. First, the prediction performance can be further boosted. In this paper, we adopt an end-to-end learning schema that trains the parameters of a neural network from scratch. However, a more efficient way in training is to 'warm-up' the network with pre-trained parameters. The pre-training can be done in a lot of ways, and some of which include the use of transfer learning algorithms. Thus, how to transfer richer information into the NeuCredit model to further improve its performance is an interesting direction. Second, in this paper, we propose a deep learning method to break down the credit risk into its determinants. Is there any other determinants can be incorporated into this framework? If there is, then how to? This is also an interesting problem to enrich the interpretability of the NeuCredit model. Third, as presented in this study, the NeuCredit model can output the predicted values of the determinants of credit risk. Therefore, they should not only be used for understanding the source of risk, but also be used for risk management. For example, based on the predictions of ability and willingness to repay, how to build more accurate models to fulfill tasks like debt collection or credit extension is also of great importance.

\clearpage

\section*{Appendix A. Derivation of the Discounting Factor} \label{app:discounting}
\addcontentsline{toc}{section}{Appendix A}

In Tva-LSTM, we assume that each element of the mapped cell memory $\bm{C}_{t-1}$ is changing at a distinct rate $R$ every unit of time during the time interval $\Delta t$. The changing rates for all the elements are denoted by a matrix $\bm{W}_R$ that has the same size as $\bm{C}_{t-1}$. Therefore, the new cell memory after $\Delta t$ is
\begin{equation}
\label{eq:interest_1}
    \bm{C}_{t-1}^D=\bm{C}_{t-1}\odot(1+\bm{W}_R)^{\Delta t}.
\end{equation}

If this change is continuous during $\Delta t$, that is $\bm{C}_{t-1}$ decays/grows $k$ times in every unit of time and $k\to+\infty$, we have
\begin{equation}
\label{eq:interest_2}
    \bm{C}_{t-1}^D=\bm{C}_{t-1}\odot\lim_{k\to +\infty}[(1+\frac{\bm{W}_R}{k})^k]^{\Delta t}.
\end{equation}

According to the definition of Euler's number $e=\lim_{n\to+\infty}(1+\frac{1}{n})^n$, Equation (\ref{eq:interest_2}) can be simplified to
\begin{equation}
    \bm{C}_{t-1}^D=\bm{C}_{t-1}\odot e^{\bm{W}_R * \Delta t},
\end{equation}
where $e^{\bm{W}_R * \Delta t}$ is regarded as a discounting factor of the mapped cell memory $\bm{C}_{t-1}$. Based on that, we introduce basic changing rates for $\bm{C}_{t-1}$ by setting up a bias matrix $\bm{B}_R$, which allows changing of the cell memory even when $\Delta t=0$. Also, an activation function is used to add non-linearity. In summary, the discounting factor becomes the one we employed in the Tva-LSTM recurrent unit:
\begin{equation}
    \bm{D}_{t-1}=e^{\operatorname{tanh}(\bm{W}_R * \Delta t + \bm{B}_R)}.
\end{equation}

\newpage
\section*{Appendix B. Generation of the Synthetic Dataset} \label{app:synthetic}
\addcontentsline{toc}{section}{Appendix B}

The synthetic dataset contains 10,000 sequences with length of 50 for each sequence. Here, we denote a sequence as $\{(\bm{x}_t,y_t)|t=1,2,...,50\}$. Each data point possesses 106 features, i.e., $\bm{x}_t=[x_{t,1};x_{t,2};...;x_{t,106}]$. $x_{t,1}$ is the time interval between data point $(\bm{x}_t,y_t)$ and $(\bm{x}_{t-1},y_{t-1})$. The value of $x_{t,1}$ is 0 when $t=1$ and is sampled from $\mathcal{U}(0,10)$ otherwise. Other features are sampled as
\begin{equation*}
    x_{t,2},...,x_{t,106}\sim\mathcal{U}(-1,1)
\end{equation*}

In generating $y_t$, only five features are involved while other 100 features are considered as noise. The computation is in the following equation, where $\mathcal{I}(\cdot)$ is the indicator function and $\sigma(\cdot)$ is the sigmoid function.
\begin{equation*}
    y_t = \mathcal{I}(\sigma(\operatorname{sin}(2x_{t,2}+x_{t,3})+3x_{t,4}x_{t,5}-x_{t,6}^3)\geq0.5)
\end{equation*}

To simulate the time-series dependencies of label and features, we recurrently generate $\bm{x}_{t+1}^p=[x_{t+1,2};...;x_{t+1,6}]$ from the time interval $x_{t+1,1}$ and the features at the previous time-stamp $\bm{x}_{t}^p=[x_{t,2};...;x_{t,6}]$. Specifically, the transformation is done using formulas below. Note that the transformation parameters $\{\bm{W}^1_{5\times5},\bm{b}^1_{5\times1},\bm{w}^2_{5\times1},\bm{b}^2_{5\times1}\}$ are not varying with time.
\begin{equation*}
\begin{aligned}
    & \bm{W}^1_{5\times5},\bm{b}^1_{5\times1},\bm{w}^2_{5\times1},\bm{b}^2_{5\times1}\sim\mathcal{U}(-1,1) \\
    & \bm{h}^p=\operatorname{tanh}(\bm{W}^1_{5\times5} \bm{x}_{t}^p + \bm{b}^1_{5\times1}) \\
    & \bm{x}_{t+1}^p = e^{(\bm{w}^2_{5\times1} x_{t+1,1} + \bm{b}^2_{5\times1})} \odot \bm{h}^p
\end{aligned}
\end{equation*}

\newpage
\section*{Appendix C. Notations for Main Variables} \label{app:notion}
\addcontentsline{toc}{section}{Appendix C}

\begin{table}[h]
	\centering
	\caption{Summary of Main Notations, in the Order of Appearance}
	\label{tab:notion_1}
	\begin{tabular}{ccp{7.7cm}c}
		\toprule
		Section & Notation & \multicolumn{1}{c}{Description} & Range/Shape \\
		\midrule
		\multirow{15}{*}{Input Definition}
		\\[-1em] & $L$ & a set of loan vectors that forms a temporal loan sequence &  \\
		\\[-1em] & $T$ & the length of a temporal loan sequence & $\mathbb{N}^+$ \\
		\\[-1em] & $\bm{l}_i$ & the vector that contains the loan features for loan $i$ in $L$ & $(d_l,1)$ \\
		\\[-1em] & $O$ & a set of order sub-sequences for loans in $L$ &  \\
		\\[-1em] & $O_i$ & the set of order vectors that forms a sub-sequence for loan $i$ &  \\
		\\[-1em] & $\bm{o}_{i,j}$ & the vector that contains the order features for order $j$ in sub-sequence $O_i$ & $(d_o,1)$ \\
		\\[-1em] & $S$ & a set of session sub-sequences for loans in $L$ &  \\
		\\[-1em] & $S_i$ & the set of session vectors that forms a sub-sequence for loan $i$ &  \\
		\\[-1em] & $\bm{s}_{i,j}$ & the vector that contains the session features for session $j$ in sub-sequence $S_i$ & $(d_s,1)$ \\
		\midrule
		\multirow{11}{*}{Sequence Encoding}
		\\[-1em] & $\Delta t$ & the time interval between the current time-stamp and the previous time-stamp & $\mathbb{R}^+$ \\
		\\[-1em] & $\bm{x}_t$ & the input vector at time-stamp $t$ & $(d_x,1)$ \\
		\\[-1em] & $\bm{h}_t$ & the hidden state of LSTM at time-stamp $t$ & $(d_h,1)$ \\
		\\[-1em] & $\bm{c}_t$ & the cell memory of LSTM at time-stamp $t$ & $(d_h,1)$ \\
		\\[-1em] & $\bm{W}_{\{i/f/o/c\}}$ & the trainable kernel matrices of LSTM & $(d_h,d_x+1)$ \\
		\\[-1em] & $\bm{U}_{\{i/f/o/c\}}$ & the trainable recurrent matrices of LSTM & $(d_h,d_h)$ \\
		\\[-1em] & $\bm{b}_{\{i/f/o/c\}}$ & the trainable bias vectors of LSTM & $(d_h,1)$ \\
		\\[-1em] & $\{\bm{i/f/o/\tilde{c}}\}_t$ & the vectors denoting the input, forget, output gates and the candidate memory of LSTM at time-stamp $t$ & $(d_h,1)$ \\
		\bottomrule
	\end{tabular}
\end{table}
\newpage

\begin{table}[h]
	\centering
	\caption{Summary of Main Notations, in the Order of Appearance (Continued)}
	\label{tab:notion_2}
	\begin{tabular}{ccp{7cm}c}
		\toprule
		Section & Notation & \multicolumn{1}{c}{Description} & Range/Shape \\
		\midrule
		\multirow{31}{*}{Sequence Encoding}
		\\[-1em] & $\bm{c}_{t-1}^S$ & the short-term memory of T-LSTM at time-stamp $t-1$ & $(d_h,1)$ \\
		\\[-1em] & $\bm{c}_{t-1}^L$ & the long-term memory of T-LSTM at time-stamp $t-1$ & $(d_h,1)$ \\
		\\[-1em] & $\bm{c}_{t-1}^{S'}$ & the discounted short-term memory of T-LSTM at time-stamp $t-1$ & $(d_h,1)$ \\
		\\[-1em] & $\bm{W}_D$ & the trainable decomposition matrix of T-LSTM & $(d_h,d_h)$ \\
		\\[-1em] & $\bm{b}_D$ & the trainable decomposition bias vector of T-LSTM & $(d_h,1)$ \\
		\\[-1em] & $\bm{C}_{t-1}$ & the mapped cell memory of Tva-LSTM & $(d_h,d_m)$ \\
		\\[-1em] & $\bm{D}_{t-1}$ & the discounting matrix of Tva-LSTM & $(d_h,d_m)$ \\
		\\[-1em] & $\bm{C}_{t-1}^D$ & the discounted mapped cell memory of Tva-LSTM & $(d_h,d_m)$ \\
		\\[-1em] & $\bm{w}_H$ & the trainable vector of Tva-LSTM that maps the cell memory to a high-dimensional space & $(1,d_m)$ \\
		\\[-1em] & $\bm{B}_H$ & the trainable bias matrix of Tva-LSTM in a high-dimensional space & $(d_h,d_m)$ \\
		\\[-1em] & $\bm{W}_R$ & the trainable matrix of Tva-LSTM that initializes the discounting factors & $(d_h,d_m)$ \\
		\\[-1em] & $\bm{B}_R$ & the trainable bias matrix of Tva-LSTM that initializes the discounting factors & $(d_h,d_m)$ \\
		\\[-1em] & $\bm{B}_D$ & the trainable bias matrix of Tva-LSTM for discounting & $(d_h,d_m)$ \\
		\\[-1em] & $\bm{w}_L$ & the trainable vector of Tva-LSTM that maps the discounted mapped cell memory back to a low-dimensional space & $(d_m,1)$ \\
		\\[-1em] & $\bm{b}_L$ & the trainable bias vector of Tva-LSTM in a low-dimensional space & $(d_m,1)$ \\
		\\[-1em] & $\bm{c}'_{t-1}$ & the new cell memory of T-LSTM or Tva-LSTM at time-stamp $t-1$ & $(d_h,1)$ \\
		\bottomrule
	\end{tabular}
\end{table}
\newpage

\begin{table}[h]
	\centering
	\caption{Summary of Main Notations, in the Order of Appearance (Continued)}
	\label{tab:notion_3}
	\begin{tabular}{ccp{7.7cm}c}
		\toprule
		Section & Notation & \multicolumn{1}{c}{Description} & Range/Shape \\
		\midrule
		\multirow{12}{*}{Multi-view Fusion}
		\\[-1em] & $\bm{h}^{o}_{i,|O_i|}$ & the final hidden state of the Tva-LSTM for order sub-sequence $O_i$ & $(d_{ho},1)$ \\
		\\[-1em] & $\bm{h}^{s}_{i,|S_i|}$ & the final hidden state of the Tva-LSTM for session sub-sequence $S_i$ & $(d_{hs},1)$ \\
		\\[-1em] & $\bm{W}_F$ & the trainable matrix that maps concatenated features & $(d_z,d_l+d_{ho}+d_{hs})$ \\
		\\[-1em] & $\bm{b}_F$ & the trainable bias vector in concatenated features mapping & $(d_z,1)$ \\
		\\[-1em] & $\bm{U}_{F1}$ & the trainable factor matrix for fusing $\bm{l}_i$ & $(d_z,d_l+1)$ \\
		\\[-1em] & $\bm{U}_{F2}$ & the trainable factor matrix for fusing $\bm{h}^{o}_{i,|O_i|}$ & $(d_z,d_{ho}+1)$ \\
		\\[-1em] & $\bm{U}_{F3}$ & the trainable factor matrix for fusing $\bm{h}^{s}_{i,|S_i|}$ & $(d_z,d_{hs}+1)$ \\
		\\[-1em] & $\bm{z}_i$ & the fused vector & $(d_z,1)$ \\
		\midrule
		\multirow{6}{*}{Hierarchical Network}
		\\[-1em] & $\bm{h}^{o}_{i,j}$ & the $j$-th hidden state of the Tva-LSTM for order sub-sequence $O_i$ & $(d_{ho},1)$ \\
		\\[-1em] & $\bm{h}^{s}_{i,j}$ & the $j$-th hidden state of the Tva-LSTM for session sub-sequence $S_i$ & $(d_{hs},1)$ \\
		\\[-1em] & $\bm{h}^{l}_{i}$ & the $i$-th hidden state of the Tva-LSTM for the sequence of fused vector $\bm{z}_i$ & $(d_{hl},1)$ \\
		\midrule
		\multirow{14}{*}{Conditional Loss}
		\\[-1em] & $\bm{w}_P$ & the trainable vector that maps $\bm{h}^{l}_{i}$ to one dimension for default probability prediction & $(1,d_{hl})$ \\
		\\[-1em] & $b_P$ & the trainable bias for default probability prediction & $\mathbb{R}$ \\
		\\[-1em] & $\hat{y}_i$ & the predicted default probability for loan $i$ & $[0,1]$ \\
		\\[-1em] & $\hat{P}_d$ & the predicted default probability & $[0,1]$ \\
		\\[-1em] & $y_i$ & the binary indicator for default of loan $i$ & $\{0,1\}$ \\
		\\[-1em] & $P(a)$ & the default probability when the ability to repay is $a$ & $[0,1]$ \\
		\\[-1em] & $P(w)$ & the default probability when the willingness to repay is $w$ & $[0,1]$ \\
		\\[-1em] & $P(b|a,w)$ & the default probability conditioned on $a$ and $w$ & $[0,1]$ \\
		\\[-1em] & $P_d$ & the default probability & $[0,1]$ \\
		\bottomrule
	\end{tabular}
\end{table}
\newpage

\begin{table}[h]
	\centering
	\caption{Summary of Main Notations, in the Order of Appearance (Continued)}
	\label{tab:notion_4}
	\begin{tabular}{ccp{7.7cm}c}
		\toprule
		Section & Notation & \multicolumn{1}{c}{Description} & Range/Shape \\
		\midrule
		\multirow{39}{*}{Conditional Loss}
		\\[-1em] & $\bm{W}_A$ & the trainable matrix that generates the hidden vector for the ability risk & $(d_{hl},d_{hl})$ \\
		\\[-1em] & $\bm{W}_W$ & the trainable matrix that generates the hidden vector for the willingness risk & $(d_{hl},d_{hl})$ \\
		\\[-1em] & $\bm{b}_A$ & the trainable bias vector that generates the hidden vector for the ability risk & $(d_{hl},1)$ \\
		\\[-1em] & $\bm{b}_W$ & the trainable bias vector that generates the hidden vector for the willingness risk & $(d_{hl},1)$ \\
		\\[-1em] & $\bm{h}_i^a$ & the hidden vector for the ability risk & $(d_{hl},1)$ \\
		\\[-1em] & $\bm{h}_i^w$ & the hidden vector for the willingness risk & $(d_{hl},1)$ \\
		\\[-1em] & $\bm{h}_i^b$ & the hidden vector for the behavioral risk & $(d_{hl},1)$ \\
		\\[-1em] & $\bm{w}_A$ & the trainable vector for the prediction of ability risk & $(1,d_{hl})$ \\
		\\[-1em] & $b_A$ & the trainable bias for the prediction of ability risk & $\mathbb{R}$ \\
		\\[-1em] & $\bm{w}_W$ & the trainable vector for the prediction of willingness risk & $(1,d_{hl})$ \\
		\\[-1em] & $b_W$ & the trainable bias for the prediction of willingness risk & $\mathbb{R}$ \\
		\\[-1em] & $\bm{w}_B$ & the trainable vector for the prediction of behavioral risk & $(1,d_{hl})$ \\
		\\[-1em] & $b_B$ & the trainable bias for the prediction of behavioral risk & $\mathbb{R}$ \\
		\\[-1em] & $\hat{y}^a_i$ & the predicted default probability for loan $i$ when ability is $a$ & $[0,1]$ \\
		\\[-1em] & $\hat{P}(a)$ & the predicted default probability when ability is $a$ & $[0,1]$ \\
		\\[-1em] & $\hat{y}^w_i$ & the predicted default probability for loan $i$ when willingness is $w$ & $[0,1]$ \\
		\\[-1em] & $\hat{P}(w)$ & the predicted default probability when willingness is $w$ & $[0,1]$ \\
		\\[-1em] & $\hat{y}^b_i$ & the predicted default probability for loan $i$ conditioned on $a$ and $w$ & $[0,1]$ \\
		\\[-1em] & $\hat{P}(b|a,w)$ & the predicted default probability conditioned on $a$ and $w$ & $[0,1]$ \\
		\\[-1em] & $r_i$ & the proportion of the installments of loan $i$ that the borrower has been delinquent on & $[0,1]$ \\
		\\[-1em] & $b$ & the batch size in mini-batch optimization & $\mathbb{N}^+$ \\
		\bottomrule
	\end{tabular}
\end{table}
\newpage

\end{document}